\documentclass{sigchi}

\newcommand{\final}{1} 

\toappear{\scriptsize Permission to make digital or hard copies of all or part of this work for personal or classroom use is granted without fee provided that copies are not made or distributed for profit or commercial advantage and that copies bear this notice and the full citation on the first page. Copyrights for components of this work owned by others than the author(s) must be honored. Abstracting with credit is permitted. To copy otherwise, or republish, to post on servers or to redistribute to lists, requires prior specific permission and/or a fee. Request permissions from permissions@acm.org. \\ {\emph{UIST '20, October 20--23, 2020, Virtual Event, USA} } \\
\copyright~2020 Copyright is held by the owner/author(s). Publication rights licensed to ACM. \\ ACM ISBN 978-1-4503-7514-6/20/10\ ...\$15.00.\\        https://doi.org/10.1145/3379337.3415892 }

\usepackage{balance}
\usepackage{graphics}
\usepackage[T1]{fontenc}
\usepackage{txfonts}
\usepackage{mathptmx}
\usepackage[pdflang={en-US},pdftex]{hyperref}
\usepackage{color}
\usepackage{booktabs}
\usepackage{textcomp}
\usepackage{microtype}
\usepackage{ccicons}
\usepackage{todonotes}
\usepackage{listings}
\usepackage{cleveref} 

\makeatletter
\def\url@leostyle{%
  \@ifundefined{selectfont}{
    \def\UrlFont{\sf}
  }{
    \def\UrlFont{\small\bf\ttfamily}
  }}
\makeatother
\usepackage{enumitem}
\usepackage{dblfloatfix}

\newcommand{\tool}{RealitySketch}

\def\plaintitle{\tool{}}
\def\plainauthor{Ryo Suzuki}

\def\plainkeywords{}

\definecolor{SithColor}{rgb}{0.7,0,0} 
\definecolor{ConsularColor}{rgb}{0,0.4,0} 
\definecolor{GuardianColor}{rgb}{0,0,0.8} 
\definecolor{PorkChopExpressColor}{rgb}{1,0.65,0} 

\newcommand {\changes}[1]{{\color{purple}{#1}\normalfont}}
\newcommand {\ryo}[1]{{\color{purple}{\bf Ryo: #1}}}
\newcommand {\rubaiat}[1]{{\color{GuardianColor}\bf{Rubaiat: #1}\normalfont}}
\newcommand {\liyi}[1]{{\color{SithColor}\bf{Li-Yi: #1}\normalfont}}
\newcommand {\daniel}[1]{{\color{ConsularColor}\bf{Daniel: #1}\normalfont}}
\newcommand {\wil}[1]{{\color{PorkChopExpressColor}\bf{WL: #1}\normalfont}}
\newcommand {\warning}[1]{{\color{purple}\bf{[#1]}\normalfont}}
\newcommand{\note}[1]{{\it\color{blue} #1}}
\newcommand {\nothing}[1]{}

\ifthenelse{\equal{\final}{1}}
{
\renewcommand {\changes}[1]{{#1}}
\renewcommand {\ryo}[1]{}
\renewcommand {\rubaiat}[1]{}
\renewcommand {\liyi}[1]{}
\renewcommand {\daniel}[1]{}
\renewcommand {\wil}[1]{}
\renewcommand {\warning}[1]{}
\renewcommand {\note}[1]{}
}
{}

\newenvironment{demofigure}[1]
{
\begin{figure}[#1]
\vspace{-0.2cm}
}
{
\vspace{-0.6cm}
\end{figure}
}

\def\pprw{8.5in}
\def\pprh{11in}

\definecolor{linkColor}{RGB}{6,125,233}
\hypersetup{%
  pdftitle={\plaintitle},
  pdfauthor={\plainauthor},
  pdfkeywords={\plainkeywords},
  pdfdisplaydoctitle=true, 
  bookmarksnumbered,
  pdfstartview={FitH},
  colorlinks,
  citecolor=black,
  filecolor=black,
  linkcolor=black,
  urlcolor=linkColor,
  breaklinks=true,
  hypertexnames=false
}

\begin{document}

\title{
\tool{}: \changes{Embedding Responsive Graphics and Visualizations in AR through Dynamic Sketching}
\vspace{-0.3cm}
}

\newcommand{\authorsep}{,}
\renewcommand{\authorsep}{\hspace{0.5em}}
\newcommand{\addrsep}{\hspace{2em}}
\numberofauthors{1}
\author{
  \alignauthor{
    Ryo Suzuki$^{1,2,3}$\authorsep
    Rubaiat Habib Kazi$^2$\authorsep
    Li-Yi Wei$^2$\authorsep
    Stephen DiVerdi$^2$\authorsep
    Wilmot Li$^2$\authorsep
    Daniel Leithinger$^{3}$\\
  \affaddr{
    $^1$University of Calgary
    \addrsep
    $^2$Adobe Research
    \addrsep
    $^3$University of Colorado Boulder
  }
  \email{
    ryo.suzuki@ucalgary.ca,
    \{rhabib, lwei, diverdi, wilmotli\}@adobe.com,
    daniel.leithinger@colorado.edu
  }
}\\
}

\teaser{
\vspace{-0.5cm}
\centering
\includegraphics[width=1\textwidth]{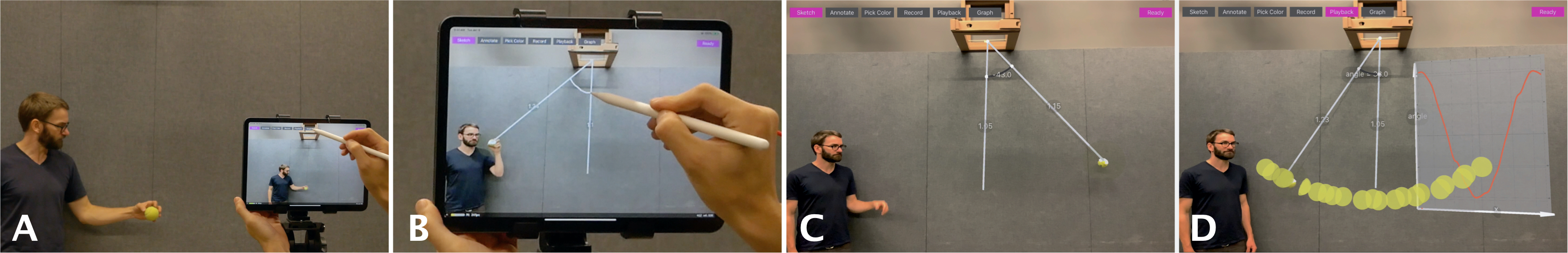}
\caption{%
\tool{} enables the user to draw and visualize physical phenomena like a pendulum's motion though real-time sketching: A) Select the ball to track, then draw a line and bind it with the physical ball. B) Draw a vertical line and an arc to parameterize the angle between the tracked ball and the center line. C) The sketched elements dynamically respond when the pendulum swings. D) Responsive graphics record and visualize the motion of the pendulum.
}
\label{fig:cover}
\vspace{-0.3cm}
}

\maketitle

\begin{abstract}
We present RealitySketch, an augmented reality interface for sketching interactive graphics and visualizations. In recent years, an increasing number of AR sketching tools enable users to draw and embed sketches in the real world. However, with the current tools, sketched contents are inherently {\it static}, floating in mid air without responding to the real world. \changes{This paper introduces a new way to embed {\it dynamic} and {\it responsive} graphics in the real world.} In RealitySketch, the user draws graphical elements on a mobile AR screen and binds them with physical objects in real-time and improvisational ways, so that the sketched elements dynamically move with the corresponding physical motion. \changes{The user can also quickly visualize and analyze real-world phenomena through responsive graph plots or interactive visualizations. 
This paper contributes to a set of interaction techniques that enable capturing, parameterizing, and visualizing real-world motion without pre-defined programs and configurations. 
Finally, we demonstrate our tool with several application scenarios, including physics education, sports training, and in-situ tangible interfaces.}
\end{abstract}



\section{Introduction}

Over the last decades, interactive and dynamic sketching has been one of the central themes in human-computer interaction (HCI) research~\cite{igarashi2007interactive, landay1996silk, landay1995interactive, mynatt1999flatland}. 
Since Sutherland first demonstrated the power of interactive sketching for computer-aided design~\cite{sutherland1964sketchpad}, HCI researchers have explored ways to sketch {\it dynamic contents} that can interactively respond to user input~\cite{kazi2014kitty, kazi2014draco, kazi2016motion, perlin2018chalktalk}, thus enabling us to think and communicate through a dynamic visual medium.
The applications of such tools are vast, including mathematics and physics education~\cite{laviola2004mathpad, scott2013physink}, animated art creation~\cite{kazi2014kitty, kazi2014draco, kazi2016motion, xing2016energy}, and interactive data visualization~\cite{brosz2013transmogrification, chao2010napkinvis, lee2013sketchstory, liu2018data, victor2013drawing, Xia:2018:DataInk}.
Through these research outcomes, we have now developed a rich vocabulary of \textit{dynamic sketching} techniques to fluidly create interactive, animated contents in real-time.

With the advent of augmented and mixed reality interfaces~\cite{milgram1994taxonomy}, we now have a unique opportunity to expand such practices beyond screen-based interactions towards {\it reality-based} interactions~\cite{ishii1997tangible, jacob2008reality, wellner1993interacting}.
In fact, there is an increasing number of tools that provide sketching interfaces for augmented reality, from commercial products like Just a Line~\cite{just-a-line}, Vuforia Chalk~\cite{vuforia-chalk}, and DoodleLens~\cite{doodlelens} to research projects like SymbiosisSketch~\cite{arora:2018:symbiosissketch} and Mobi3DSketch~\cite{kwan:2019:mobi3dsketch}.
These tools allow users to sketch digital elements and embed them in the real world. However, a key limitation is that the sketched content is {\it static} --- it does not respond, change, and animate based on user actions or real-world phenomena. 

What if, instead, these sketched elements could dynamically respond when the real world changes? 
For example, imagine a line sketched onto a physical pendulum that moves as the pendulum swings (Figure~\ref{fig:cover} A-B). This capability would allow us to directly manipulate the sketch through tangible interactions (Figure~\ref{fig:cover} C) or capture and visualize the pendulum motion to understand the underlying phenomenon (Figure~\ref{fig:cover} D).
Such responsive and embedded sketching would enable a new way of seeing, understanding, and communicating ideas in much richer ways~\cite{perlin2016future, victor2014humane, victor2014seeing}, by bridging the gap between abstract theories and concrete real-world experiences~\cite{kay2005squeak}.

As a first step toward this goal, we present \tool{}, an end-user sketching tool to support real-time creation and sharing of embedded interactive graphics and visualization in AR. 
To create graphics, the user sketches on a phone or tablet screen, which embeds interactive visualizations into a scene. The sketched elements can be bound to physical objects such that they respond dynamically as the real world changes. 
%
%

Our approach proposes the following four-step workflow: 
1) {\it object tracking}: the user specifies a visual entity (e.g., a physical object, a skeletal joint) to track in the real-world scene;
2) {\it parameterization}: the user parameterizes tracked entities by drawing lines or arcs that define specific variables of interest;
3) {\it parameter binding}: the user binds these variables to the graphical properties of sketched elements (e.g., length, angle, etc.) to define their dynamic behavior as the real-world variables change;
4) {\it visualization}: the user can also interact, analyze, and visualize real-world movements through several visualization effects. 
We contextualize our approach in the larger design space of dynamic graphics authoring approaches.
The main contribution of this paper is a set of interaction techniques that enable these steps without {\it pre-defined} programs and configurations --- rather, the system lets the user perform in {\it real-time} and {\it improvisational} ways. 

We demonstrate applications of \tool{} across four domains:
augmented physics experiments, mathematical concept explorations, sports and exercise assistants, and creation of improvised tangible UI elements.
These examples illustrate how improvisational visualization can enrich interactive experiences and understanding in various usage scenarios.
Based on a preliminary evaluation and an expert interview, we discuss the benefits and limitations of the current prototype.
\changes{Our prototype runs on mobile AR consumer devices that support ARKit (iPhones and iPads), which is important for adoption in an educational context. However, we believe that in the future, our proposed interactions and use cases can also be adapted for head-worn devices.}

In summary, this paper makes the following contributions:
\begin{enumerate}[itemsep=-1mm]
\item \tool{}, a \changes{\textit{dynamic sketching}} system for real-time authoring of embedded and responsive visualizations in AR, contextualized in the larger design space of dynamic graphics authoring approaches.
\item A set of sketching interaction techniques to parameterize the real-world and bind it with embedded graphics.
\item Application scenarios where the users can create and view in-situ visualizations by capturing physical phenomena, such as for classroom experiments\nothing{ support} and interactive concept explanations.
\end{enumerate}

\section{Related Work}

\subsection{Sketching Interfaces}
In literature of HCI, there is a long history of sketching interface research~\cite{bae2008ilovesketch, gross1996ambiguous, igarashi2007interactive, igarashi2006teddy, landay2001sketching, mynatt1999flatland, sutherland1964sketchpad}. 
Since the space of sketching interface research is vast, here, we focus on two specific areas of research that are highly relevant.

\subsubsection{Sketching Interfaces for VR/AR}
In recent years, a growing number of tools support sketching in VR and AR~\cite{thoravi2019tutorivr}.
For example, Just a Line~\cite{just-a-line}, TiltBrush~\cite{tiltbrush}, Gravity Sketch~\cite{gravity-sketch}, and Vuforia Chalk AR~\cite{vuforia-chalk} are common sketching software available in the market.
These tools allow users to sketch objects in which they can walk around~\cite{tiltbrush, gravity-sketch, wesche2001freedrawer} or annotate sketches onto a video see-through AR view of the real-world environment~\cite{vuforia-chalk}.
However, due to the challenges of mid-air 3D sketching~\cite{arora2017experimental, barrera2017multiplanes}, researchers have leveraged real-world environments and physical objects as contextual guidelines to sketch virtual objects~\cite{schkolne2001surface}.
For example, a physical surface can serve as a geometric constraint for sketching in an immersive environment \changes{(e.g., SymbiosisSketch~\cite{arora:2018:symbiosissketch}, VRSketchIn~\cite{drey2020vrsketchin}, PintAR~\cite{gasques2019you})}, which gives a useful guidance for 3D sketches~\cite{arora2017experimental}.
Also, previous work utilizes real-world geometry as contextual guidelines for snapping elements \changes{(e.g., Mobi3DSketch~\cite{kwan:2019:mobi3dsketch}, SnapToReality~\cite{ nuernberger2016snaptoreality})} or making 3D shapes \changes{(e.g., SweepCanvas~\cite{li:2017:sweepcanvas}, SketchingWithHands~\cite{kim2016sketchingwithhands})}.

However, most of these current sketching tools only support {\it static} drawings --- once sketched, the sketched graphics do not animate or interact with real-world \changes{(Figure~\ref{fig:design-space} bottom right)}.
DoodleLens~\cite{doodlelens} explores the creation of animated drawings with flip-book effects of multiple drawings, but it is not interactive, in the way the sketched elements do not respond to a user's interaction. 
Video prototyping tools like Pronto~\cite{leiva:2020:pronto} also let the user animate embedded sketches, but they only mimic the interaction for prototyping purposes. ProtoAR~\cite{nebeling2018protoar} and 360Proto~\cite{nebeling2019360proto} facilitates the creation of AR and VR experiences with paper sketches, but the sketches themselves are static.

If the sketched graphics become responsive, we can leverage our rich tangible and gestural manipulation capability to interact with the virtual world.
For example, in the context of animation authoring, real-world objects provide useful and expressive ways to create dynamic motion by tangible interactions~\cite{barnes:2008:video, glauser2016rig, held:2012:3dpuppetry}.
Human-motion can be also used as real-time inputs for making performance-driven animations~\cite{Gambaretto:2014:RAC, saquib2019interactive}.
However, these tools often require tedious pre-defined configurations to map between the virtual content and its physical reference (e.g., physical objects, a rig of the body, segmentation of the face), which requires a lot of preparations in advance.
It is important to develop more real-time and improvisational ways to specify the behavior because such a practice can preserve a user's natural workflow~\cite{kazi2012vignette} and let the user focus more on creation and communication~\cite{perlin2016future, victor2012stop, victor2013drawing}.

\subsubsection{Sketching Dynamic and Responsive Graphics}
HCI researchers have also extensively explored sketching interfaces for dynamic and responsive graphics.
For example, MathPad2~\cite{laviola2004mathpad} and PhysInk~\cite{scott2013physink} use hand-drawn sketches to construct an interactive mathematical and physics simulation.
In contrast to pre-programmed simulation \changes{(e.g., PhET~\cite{perkins2006phet, wieman2008phet}, Explorable Explanations~\cite{explorables, victor2011explorable})}, such improvisational sketching enables more flexible applications and encourages natural interaction for human-to-human communication~\cite{lee2013sketchstory}.
This dynamic sketching idea has been implemented for other application domains, such as animation authoring 
\changes{(e.g., Kitty~\cite{kazi2014kitty}, Draco~\cite{kazi2014draco}, Apparatus~\cite{schachman2015apparatus}, Motion Amplifiers~\cite{kazi2016motion}, and other tools~\cite{victor2012stop, xing2016energy, zhu2011sketch})} 
and dynamic data visualizations 
\changes{(e.g., SketchStory~\cite{lee2013sketchstory}, NpakinVis~\cite{chao2010napkinvis}, Data Illustrator~\cite{liu2018data}, Transmogrification~\cite{brosz2013transmogrification}, and Drawing Dynamic Visualizations~\cite{victor2013drawing})}.

\begin{figure}[t!]
\centering
\includegraphics[width=1\linewidth]{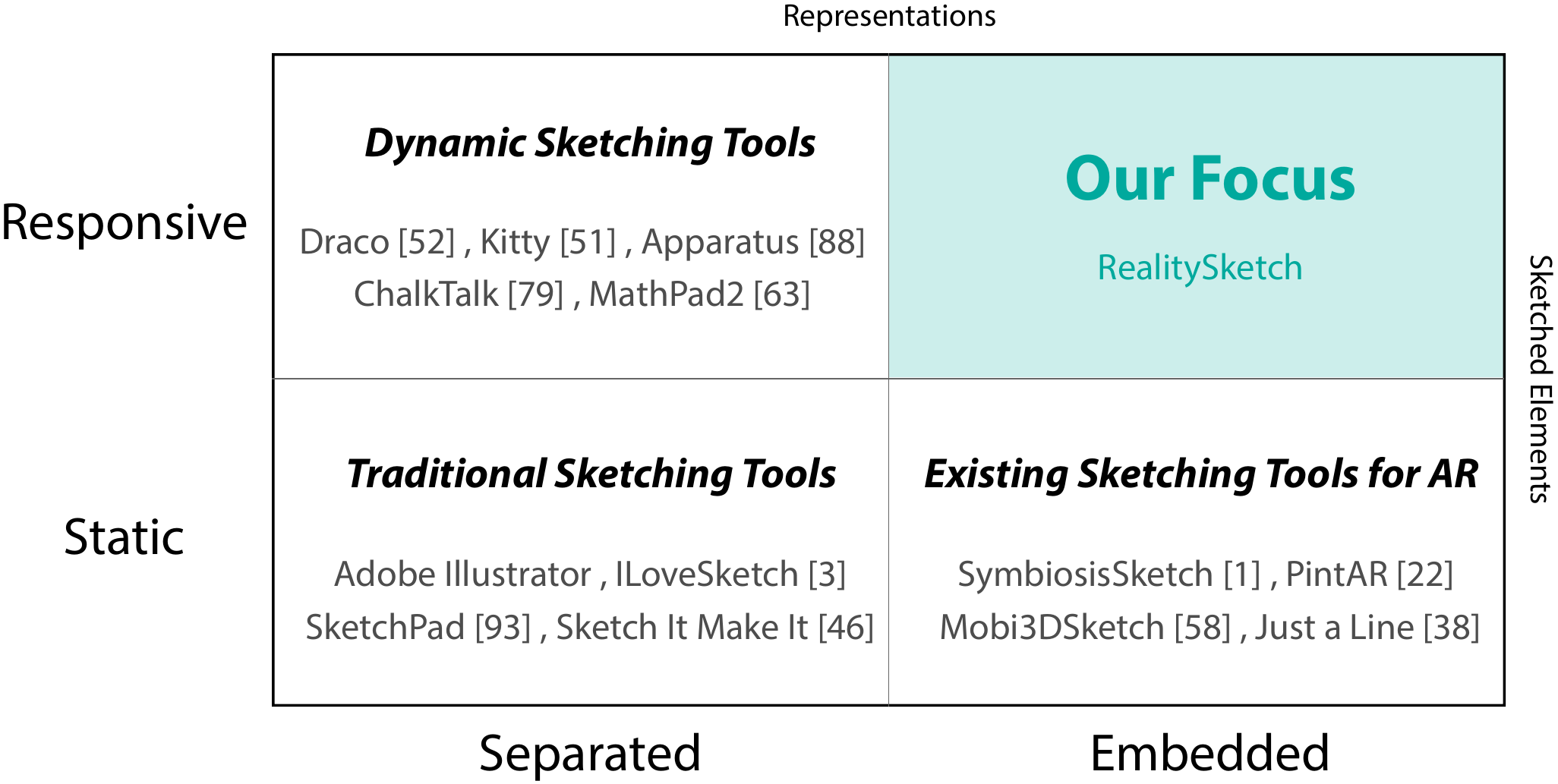}
\caption{Design space of \changes{dynamic sketching} interfaces: The horizontal axis shows whether the representation is embedded (graphics are embedded in the real world) or separated (graphics are isolated from the real world). The vertical axis shows whether sketched elements are dynamic/responsive (sketched elements can dynamically move and respond) or static (sketched elements do not change). Our focus  is on embedded and responsive sketching (top right).}
~\label{fig:design-space}
\vspace{-0.8cm}
\end{figure}

While most of these interfaces are for {\it screen-based} applications, {\it augmented reality (AR) based} interfaces have great potential to further enhance interactive experiences and natural communication in the real world.
For example, ChalkTalk AR~\cite{perlin2018chalktalk, perlin2018chalktalkvrar} enables sketching dynamic graphics on a see-through display so that the user can present and brainstorm ideas through face-to-face communication.
However, these sketched elements do not {\it respond to} or {\it interact with} the real-world environments. \changes{In other words, the sketched contents are not spatially embedded nor integrated within the real world (Figure~\ref{fig:design-space} top left)}.
In contrast, this paper explore {\it embedded} responsive sketches that can directly interact with the real world, so that it can provide more immersive experiences for AR and MR \changes{(Figure~\ref{fig:design-space} top right)}.

In the context of tangible user interfaces, there are a few works that explore tangible interactions with sketched elements \changes{(e.g., Reactile~\cite{suzuki2018reactile})}.
Inspired by this line of work, this paper contributes to the discussion and exploration of the new sketching interactions to support a more diverse set of applications.

\subsection{Augmented Reality and Tangible User Interfaces}
\subsubsection{Embedded Data and Concept Visualization}
Interactive graphics in AR can facilitate communication, education, storytelling, and design exploration.
For example, Saquib et al.~\shortcite{saquib2019interactive} and Liu et al.~\shortcite{Liu:2020:PPT} demonstrated how interactive and embedded graphics in video presentation can augment storytelling\nothing{ of complex ideas} and performance.
Similarly, ARMath \cite{kang2020armath} demonstrates how everyday objects can be used as tangible manipulators to learn basic arithmetics.
Data visualization is another promising area, as the user can directly see the data that is associated with physical objects~\cite{raskar2004rfig}.
Willett et al.~\cite{willett2016embedded} define such a representation as {\it embedded visualizations}, in which the visualization is tightly coupled with its physical referent.

Traditionally, such experiences are only available after the dedicated post-production process (e.g., educational videos like~\cite{hidden, sports-explainer, byjus, nova}), but augmented and mixed reality interfaces promise an {\it interactive} experience for such concept and data visualizations.
For example, in the context of physics education, existing works explore ways to visualize a range of invisible phenomena, \changes{such as wind flow (e.g., Urp~\cite{underkoffler1999urp}), optical paths (e.g., HOBIT~\cite{furio2017hobit}), terrain simulation (e.g., Illuminating Clay~\cite{piper2002illuminating}), electromagnetic flow (e.g.,~\cite{radu2019can}), and current flow (e.g., VirtualComponent~\cite{kim2019virtualcomponent}, ConductAR~\cite{narumi2015conductar})} in the real world.
These interfaces can facilitate understanding of a complex idea~\cite{dunleavy2014augmented}, as viewers can directly observe the hidden phenomenon, instead of imagining it~\cite{victor2014seeing}.
Beyond education, embedded and interactive visualizations are also promising in application domains like collaborative discussions~\cite{dillenbourg2013design, kasahara2012second}, live storytelling~\cite{saquib2019interactive}, data visualization~\cite{chen2019marvist}, and sports training~\cite{chen2018computer, homecourt, sousa2016augmented}. 

Currently, however, the interactions with these systems are {\it pre-programmed} for specific applications by their developers, and do not allow non-technical users to adapt them to different contexts or create custom experiences by themselves.
With the real-time and improvisational authoring introduced in this paper, we believe end users can easily create, manipulate and share these contents in more engaging and expressive ways, across different application domains.

\subsubsection{Object Manipulation through Spatial Tablet Interactions}
Spatially aware tablet computers can provide an interface to manipulate virtual and real-world objects. Examples include T(Ether)~\cite{lakatos2014t}, where users point and drag virtual objects with a spatially tracked tablet computer. ExTouch~\cite{kasahara2013extouch} introduces a similar interaction technique to manipulate robots at a distance. When tapping on a robot shown in an AR video feed, users can drag it to a new target location. By sketching connections between IoT devices, users of Smarter Objects~\cite{heun2013smarter} and Reality Editor~\cite{heun2013reality} can program their functionality.
Our work extends these ideas of interacting with real-world objects through tablet-based AR. But instead of programming the objects they serve as an input for sketched responsive graphics.


\begin{figure*}[h!]
\vspace{-0.2cm}
\centering
\includegraphics[width=0.95\linewidth]{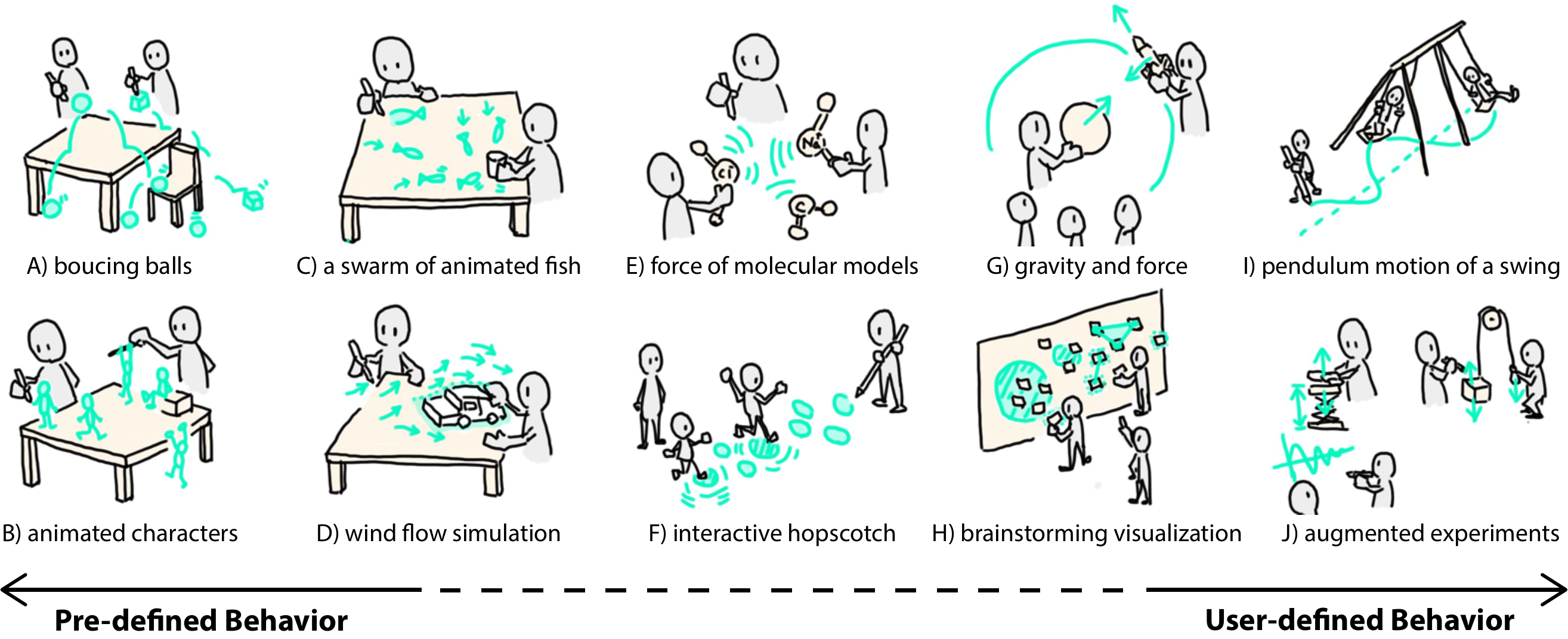}
\caption{Potential examples of embedded and responsive sketches (green elements represent objects in the virtual world, and black elements represent objects in the real world).
}
~\label{fig:pre-defined-user-defined}
\vspace{-0.6cm}
\end{figure*}

\section{Sketching Embedded and Responsive Graphics}
\label{sec:design}

\subsection{Definition of Embedded and Responsive Graphics}
The goal of this paper is to provide a way to embed dynamic and responsive graphics through \changes{dynamic sketching}.
To better understand the scope, we first define the terminology:
\begin{description}[itemsep=-1mm]
\item[Embedded:] graphics are presented \changes{and spatially integrated within} the real-world 
\item[Responsive:] graphics change and animate based on the user's interactions
\end{description}
The important aspect of embedded and responsive graphics is that {\it graphics must interact with the real-world}.
Here, the ``real-world'' means either the environment itself, a physical phenomenon, or a user's tangible and gestural interactions. 
In the context of AR and tangible user interfaces, there are many non-sketched examples of such embedded and responsive graphics. For example, animated fish in PingPongPlus~\cite{ishii1999pingpongplus} can interact with the ball by moving towards a point where the physical ball hits, or virtual cubes in HoloDesk~\cite{hilliges2012holodesk} can interact with physical environments by rolling and bouncing on top of it. 

However, to our best of knowledge, few works, if any, investigated {\it sketching} as a way to create such interactive experiences in real-time.
Therefore, this section provides the first exploration and discussion about the interaction spectrum for authoring such embedded and responsive graphics to explore the broader design space.
By taking inspirations from the existing works in dynamic sketching of 2D drawing, we discuss and illustrate possible approaches through potential sample applications.
We believe this discussion and exploration will help the HCI community to better understand the broader design space and further explore \changes{it} to fill the gap.

\subsection{How to Make Embedded Sketches Responsive }
In general, crafting responsive and embedded graphics in the real-world can be a continuum between two approaches: pre-defined behavior and  user-defined behavior.
Since our focus is on interactive experiences, we do not discuss post-production video editing tools. 

\subsubsection{Pre-defined Behavior vs User-defined Behavior}
Pre-defined behavior refers to a behavior specification given in advance.
For example, one can think of a system that specifies all of the sketched elements to follow the law of physics, so that as long as a user draws a virtual element, it automatically falls and \changes{bounces} on the ground (Figure~\ref{fig:pre-defined-user-defined} A).
In this case, the behavior of sketched elements is pre-defined, based on the physics simulation, and the user can only control the shape of the sketches. 
Similarly, \nothing{for example, }one can imagine a sketched character that starts walking around or interacting with the physical environment (Figure~\ref{fig:pre-defined-user-defined} B).
In this case, the behavior of the sketched character should also be defined in advance (by programming or design tools), as it is hard to specify such complex behaviors in real-time.

These practices are often utilized in the screen-based sketching interfaces.
For example, PhysInk~\cite{scott2013physink} uses a physics engine and ChalkTalk~\cite{perlin2018chalktalk} leverages pre-programmed behavior to animate the sketched elements in real-time. 

On the other end of the spectrum, user-defined behavior lets the user decide how the sketched elements move, behave, and animate on the fly.
For example, consider an example of visualizing pendulum motion (Figure~\ref{fig:cover}).
In this example, the user should be able to specify how and which parameter (e.g., angle) will be visualized (Figure~\ref{fig:pre-defined-user-defined} I).
In the previous works, Apparatus~\cite{schachman2015apparatus} leverages the user-defined behavior to create interactive diagrams.
In this example, the user has full control of how it behaves, based on the user-defined constraints and parameter bindings, which is also known as constraint-based programming~\cite{borning1981programming, borning1986constraint, sutherland1964sketchpad}.
These practices are also utilized to create \nothing{an }interactive 2D animation~\cite{victor2012stop}, \changes{design exploration~\cite{lee20203d}}, and dynamic data visualization~\cite{liu2018data, victor2013drawing}, as it is useful to let the user explicitly specify how it behaves.

We can also think of a design that leverages a combination of both pre-defined and user-defined approaches~\cite{zhu2011sketch}.
For example, consider a sketched school of fish that can automatically swim on a table (Figure~\ref{fig:pre-defined-user-defined} C). The basic swimming motion can be specified by pre-defined behavior, but the user could interactively specify the target position so that the fish can continuously chase and gather to a physical object.
\changes{This can be achieved by by leveraging direct manipulation interfaces that can be paired with the pre-defined program (e.g., Sketch-n-sketch~\cite{chugh2016programmatic}, Eddie~\cite{sarracino2017user}).}

Both approaches have advantages and disadvantages.
For example, pre-defined behavior enables highly complex and expressive animation (e.g., Figure~\ref{fig:pre-defined-user-defined} A-D), but is targeted at specific applications (e.g., character animation, physics simulation, etc).
Beyond these well-known configurations, it is difficult to generalize the range of behaviors and effects.
On the other hand, user-defined behaviors can provide building blocks that let the user construct them for different purposes, thus it can be generalizable (e.g., Figure~\ref{fig:pre-defined-user-defined} G-J).
Also, the user does not need to memorize what behaviors are possible, as it is based on a set of simple interactions.
However, creating complex behaviors with low-level building blocks is challenging, particularly when the interaction is happening in real-time for communication and exploration. \changes{Thus, finding the right balance between these two approaches is dependent upon the application and user needs.}

\subsubsection{Focus of This Paper}
Given these considerations, this paper specifically focuses on {\it user-defined behavior} approach for the input specification and mapping of {\it embedded} graphics.
However, we leverage {\it pre-defined} visual outputs (e.g., graph plots) so that the user does not need to create these commonly used graphical elements from scratch in real-time.
To this end, there are the following research questions:
\begin{description}[itemsep=-1mm]
\item[Q1. Input]
What is the scope of input values? How can the user visually annotate and define the input value? 
\item[Q2. Input-Output Mapping]
How can the user map the input values to output graphical elements on the fly?
\end{description}

\nothing{
The first aspect is input.
Since the interactive behavior is triggered or associated with its input values, it is important to consider what we should take as an input.
The second aspect is input and output association.
In user-defined behavior, the interactive behavior is often described as an association between input values and the property of sketched elements, thus how to bind the different parameters to become the key.
Since the sketched graphics respond based on the change of other input values, the system must support the interactive and intuitive way to specify how these values should associate each other.} 

The input, output (graphics), and mapping space can vary across application domains and user needs. We aim to demonstrate interaction techniques for input, output, and mapping that are useful across several application domains (e.g., physics experiments, personal training) where spontaneous and interactive exploration is key to effective performances.
In this paper, we answer these research questions by exploring:
\begin{description}[itemsep=-1mm]
\item[A1. Input]
Parameterizing the real world through sketching interactions. These parameterized values can be later used for the input values of other graphical elements. Bertin \cite{Bertin:1983:SG} introduced a set of visual variables to encode information. In this paper, we focus on continuous spatial variables (e.g., position, distance, angle) as input, as they relate to physical phenomena, tangible interactions, and movements. Other forms of visual variables (e.g., shape, size, type) and gestural interactions are beyond the scope of this paper. 

\item[A2. Input-Output Mapping]
Defining a dynamic parameter value as a variable on-the-fly, so that the variable can be used to map input and output variables. 

\end{description}

For visual output, we use pre-defined graphical elements, such as dynamic sketched elements (with the combination of object tracking and parameter binding), responsive visualizations like graph plots, and visual effects like stroboscopic images by recording the motion. We chose these graphical elements due to their effectiveness and versatility to analyze and interact with spatial variables. 

\begin{figure}[h!]
\vspace{-0.2cm}
\centering
\includegraphics[width=1\linewidth]{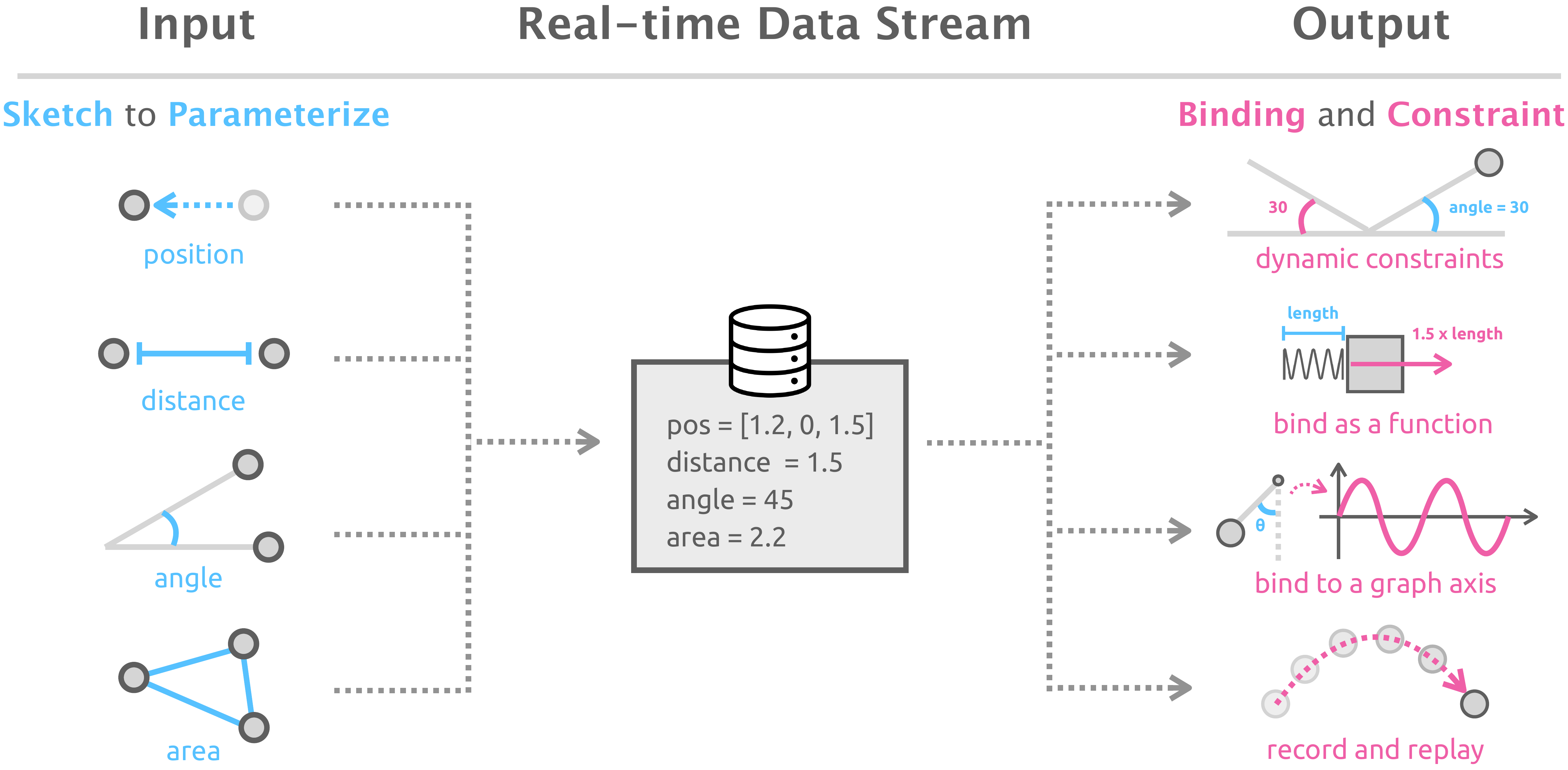}
\caption{\changes{Overview of the interaction workflow:
Sketched spatial variables (e.g.,
position, distance, angle, etc) can be used as dynamic input values. 
By binding this real-time data stream to existing graphics, the user can create responsive graphics.}
}
~\label{fig:system-overview}
\vspace{-0.6cm}
\end{figure}

\section{RealitySketch: System and Design}
\label{sec:system}

\subsection{Overview}
This section introduces \tool{}, an augmented reality interface to embed responsive graphics and visualizations through real-time \changes{dynamic sketching}.

\tool{} supports the followings operations to draw dynamic sketches based on the user-defined specification:
\begin{description}[itemsep=-1mm]
\item[1. Track] an object by specifying it on the AR screen.

\item[2. Parameterize] the real world with sketching.

\item[3. Bind] the real-world parameters into existing graphics to make the graphics responsive.

\item[4. Visualize] the real-world dynamics through responsive graph plots or recording and playing back the motion.
\end{description}




\subsection{Basic Setup}
\begin{figure}[h!]
\vspace{-0.2cm}
\centering
\includegraphics[trim={0 300 0 0},clip,width=0.9\linewidth]{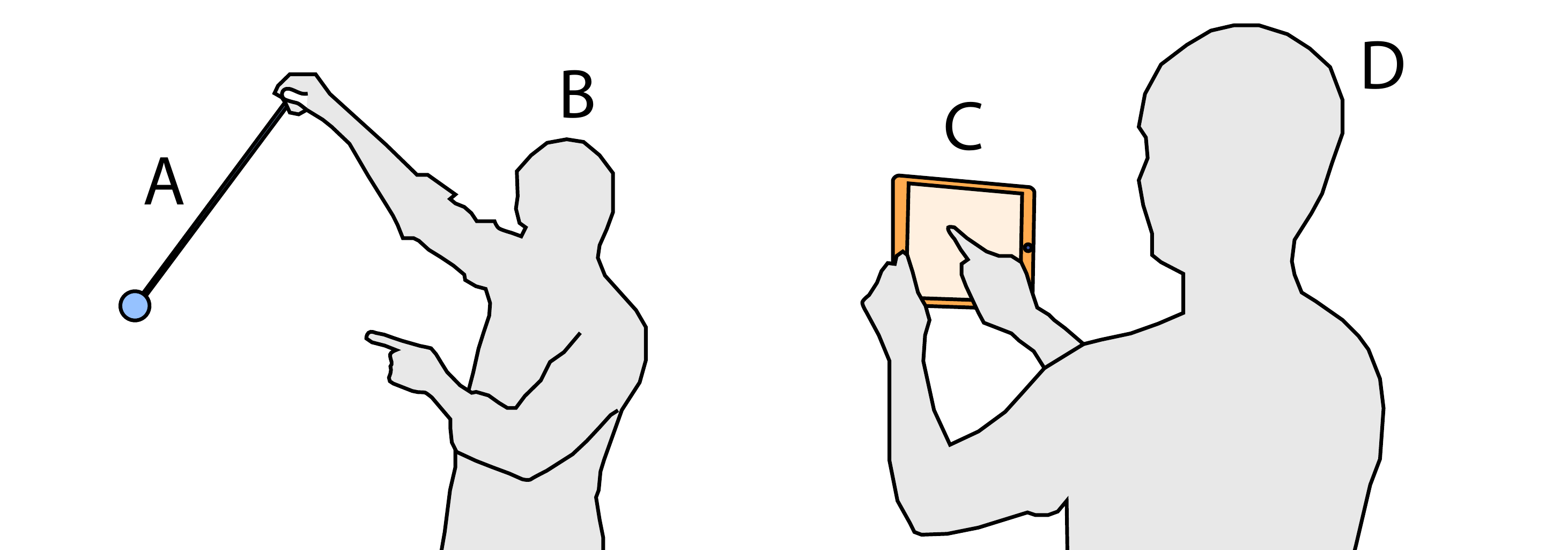}
\caption{System setup: The user (D) sketches and \changes{manipulates} on a tablet or phone (C).
\changes{The user can sketch on a real-time scene or a recorded scene.}
If necessary, the other user (B) can collaborate to move or interact with physical objects (A).
}
~\label{fig:system-setup}
\vspace{-0.6cm}
\end{figure}

\tool{} leverages mobile augmented reality (ARKit) to embed sketches in the real world.
The user sketches with a finger or a pen on a touchscreen, where the sketched elements are overlaid onto a camera view of the real world \changes{(Figure~\ref{fig:system-setup})}.

\changes{Rather than 2D sketches that are based on the device screen coordinates, all sketched elements have a 3D geometry and position in real world coordinates, anchored in 3D space. This way, the user can move the device to see from a different perspective and the sketches stay correctly anchored to the real objects.

To enable this functionality, our system leverages ARKit and SceneKit for both surface detection and object placement in the 3D scene. Therefore, the first step is to detect a horizontal or vertical surface, as all interactions (e.g., tracking an object, sketching elements, placing a graph) are based on a reference surface in a 3D scene.}
Once the surface is detected, the user can start using the system.


In \tool{}, the user can sketch on both real-time and pre-recorded scenes.
For real-time sketching, the user can just start sketching once the surface detection is finished. 
For recorded scenes, \tool{} first allows users to capture and record the video with an internal recording feature.
The recorded video is embedded with additional meta information at each time frame, such as the current camera position, the current origin point, and the position of detected surface.
Based on this recorded information, \tool{} can reconstruct the 3D scene, which is associated with the recorded camera view of each time frame.
Thus, the user can sketch, annotate, or visualize on top of the \tool{}'s recorded video, while controlling the timeline of the internal video player.
Due to these constraints, this recording feature does not work with conventional videos.






\subsection{Object Tracking}
For embedded and responsive graphics, the graphical elements need to be tightly coupled with physical objects and environments.
Thus, capturing and tracking an object is vital to make the graphics dynamically change and move.

To specify an object, the user enters the selection mode and then taps an object on the screen (Figure~\ref{fig:system-tracking}A). Once selected, our system highlights the selected object with a white contour line (Figure~\ref{fig:system-tracking}B) and starts tracking the object in the 3D scene (Figure~\ref{fig:system-tracking}C). 

\begin{demofigure}{tbh!p}
\centering
\includegraphics[width=1\linewidth]{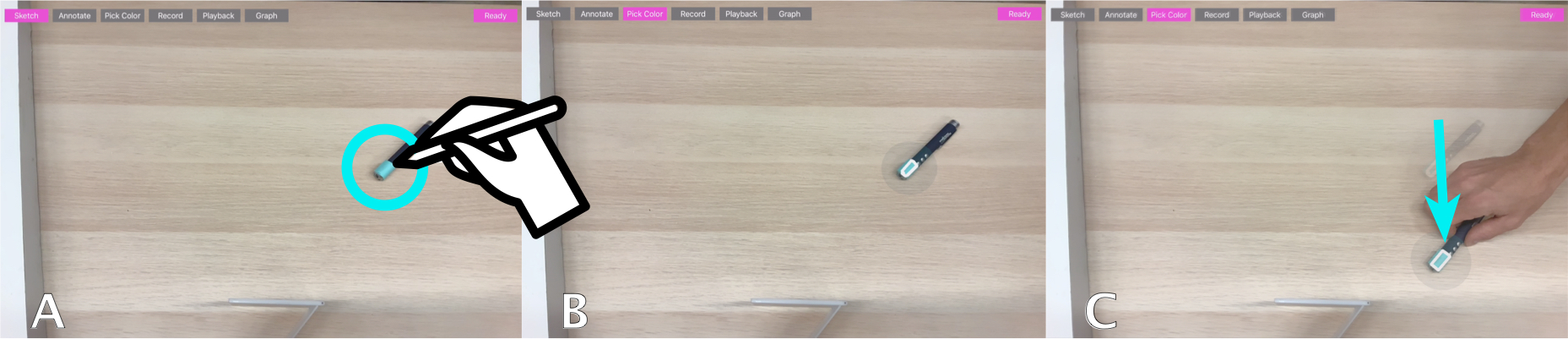}
\caption{Object tracking: Tap an object to be tracked (A), and the system starts tracking the object based on color matching (B-C).}
~\label{fig:system-tracking}
\end{demofigure}

\changes{In our current implementation, the system tracks objects based on color tracking implemented with OpenCV. When the user taps an object on the screen, the algorithm gets the HSV value at the x and y position. Then the system captures similar colors at each frame based on a certain upper and lower threshold range.}
Therefore, the tracking feature best works with distinct, solid colors.
Based on this detected mask, the system obtains the largest contour and computes a center position \changes{of the tracked object on the screen. The system then converts these 2D coordinates into the 3D world coordinates} by ray casting the 2D position onto the detected surface, assuming that the tracked object moves onto the detected horizontal or vertical surface. 
This color tracking was fast enough for most of our applications (e.g., 20-30 FPS with iPad Pro 11 inch 2018).
The user can also use the joint of the human body as additional tracked objects.
In this case, we use ARKit 3's built-in human motion tracking feature that gives us the 3D position information of the body joints.

\begin{demofigure}{h!}
\centering
\includegraphics[width=1\linewidth]{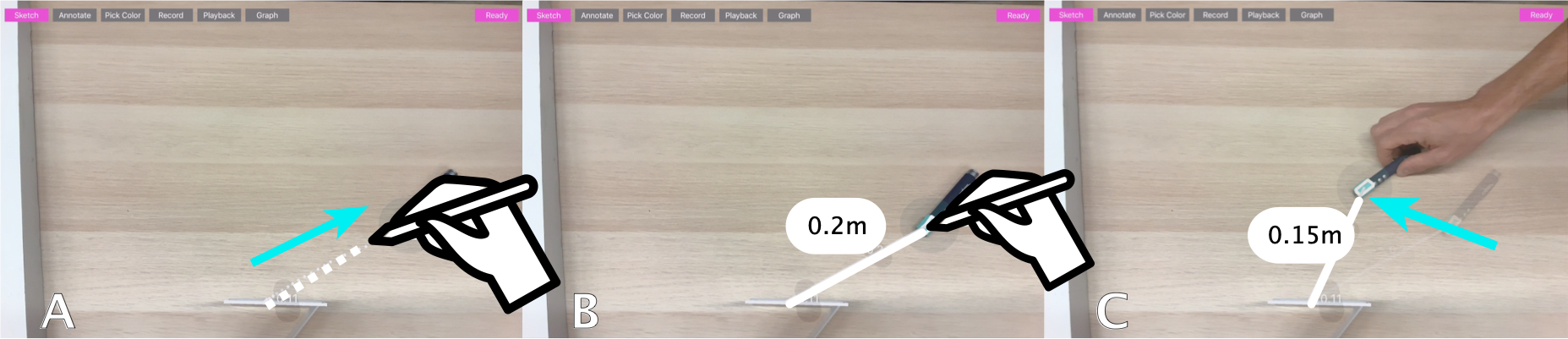}
\caption{Line segments: Sketch a line on the screen (A). If the line is attached to the tracked object (B), it becomes bound to the object (C).}
~\label{fig:system-line}
\end{demofigure}

\subsection{Parameterization using Line Segments}
\changes{Next, the user parameterizes the real world to define and capture the dynamic value of interest.}
In this paper, we specifically focus on parameterization that can be done through simple sketching interactions using {\it line segments}. 
Line segments are one of the most basic sketching elements, which are commonly leveraged in the existing interactive sketching systems (e.g., ~\cite{igarashi2007interactive}).
Line segments are useful as they work alone to parameterize a length between two points, but also can be combined to define different parameters, such as an angle between lines or an area of a closed geometric shape.
They can be also used for annotated lines or guided lines.
The user could also draw a simple 3D diagram with these line segments like~\cite{igarashi2007suggestive}, which could be used as a responsive visual output itself.

Here, we describe how this line sketching interaction can define and capture the dynamic value of interest.
First, when entering the sketching mode, the user can start drawing a line segment onto the scene (Figure~\ref{fig:system-line}A).
All the sketched lines are projected onto a 2D surface within the 3D scene.
The system takes the two end-points to create the line segment. 
This creates a variable that defines the distance between two points on the surface.
To create a \textit{dynamic line segment}, the user draws a line whose end point is attached to the selected object (Figure~\ref{fig:system-line}B-C).
As one end of this \textit{dynamic line segment} is bound to the selected object, if the user moves the object in the real world, the line segment and its parameter (e.g., distance value) will change accordingly. 
The system visually renders the line segment values using labels.

\begin{demofigure}{h!}
\centering
\includegraphics[width=1\linewidth]{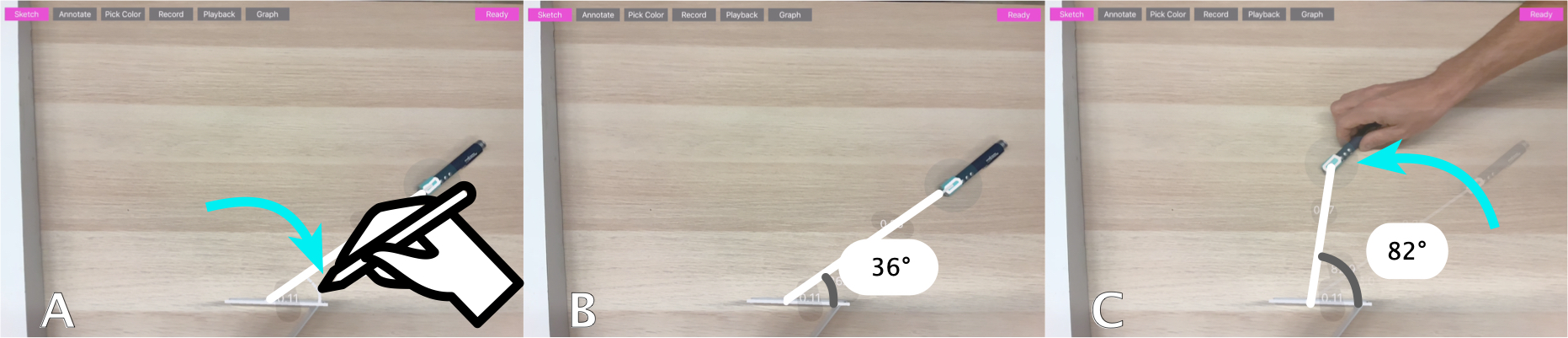}
\caption{Angle: The user can draw a path between lines (A), and the system parameterizes it as an angle (B). The angle can be dynamic if the line is bound to the object (C).}
~\label{fig:system-angle}
\end{demofigure}

\tool{} employs simple heuristics to determine the nature (e.g., static vs. dynamic, distance vs angle, free move vs constraints, etc) of the line segment.
If the start or end point of the line segment is close to an existing tracked object, then the system binds the end point to the tracked object.
Thus, for example, if the user draws a line between two tracked objects, then both ends attach to an object. In that case, the line segment captures the distance between those two objects.

The user can also define an area of a closed enclosure or an angle between multiple line segments as a variable.
For example, if the user draws a line between the end points of two different lines, then the system binds the three lines together to parameterize the enclosed area of the shape (e.g., a triangle).
On the other hand, when the user sketches a stroke between two lines (Figure~\ref{fig:system-angle}A), the system recognizes it as an angle and shows an arc between the lines (Figure~\ref{fig:system-angle}B).
To determine whether the drawn line is a distance or an angle, the system only considers the start and end points.
If both points are attached or close to the existing line segments, the user-drawn line is interpreted as an angle.
If the existing line is bound to an object, then the angle is also dynamically changed when the object is moved (Figure~\ref{fig:system-angle}C).
In a similar manner, if the two lines are close to perpendicular, the system creates a perpendicular constraint for two lines (e.g., a tracked line and the ground line in Figure~\ref{fig:applications-gravity}).

The values of these parameters (e.g., length, angle, area) are actual physical quantities in the real world, as we can measure the actual length of two points in the 3D scene using ARKit's measuring capability.
We use this value for the distance parameter of a sketched line.

\subsubsection{Naming Variables}
The user can also assign a name to an existing parameter e.g., {\it x, y, angle, length}, etc.
This entered name works as a variable that can be later used for input-output mapping. 
When the user taps the label of the dynamic parameter (Figure~\ref{fig:system-naming}A), the system shows a popup dialog to let the user define a variable name (Figure~\ref{fig:system-naming}B).
Once the user types the variable name, the system registers it as a variable and shows the name in the label (e.g., $angle = 40$ in Figure~\ref{fig:system-naming}C).

\begin{demofigure}{h!}
\centering
\includegraphics[width=1\linewidth]{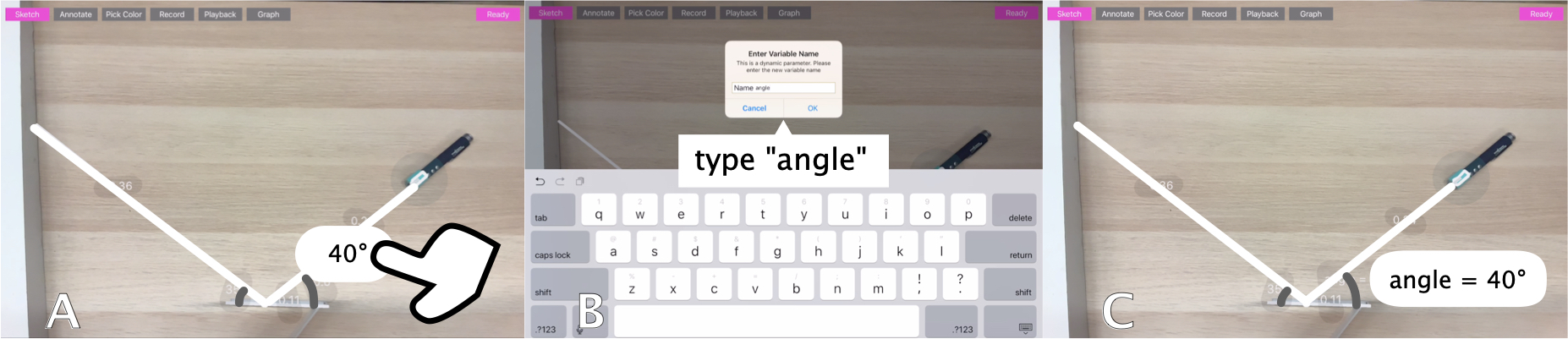}
\caption{Naming a variable: By tapping a label of a line segment (A), the user can name a variable (B-C).}
~\label{fig:system-naming}
\end{demofigure}

\subsection{Parameter Binding}


\subsubsection{Make Graphics Responsive based on Parameter Binding}
To make the existing line segments responsive, the user can bind variables between two elements.
The parameter of a \textit{static line segment} can be bound to another variable.
For example, suppose the user has a variable named $angle$ for a {\it dynamic line segment}.
When the user taps the label of another angle of the {\it static line segment} (Figure~\ref{fig:system-bind}A), \nothing{then }the system shows a popup to let the user enter the variable name.
If the entered variable name matches an existing name (e.g., $angle$ in this case), the angle of the {\it static line segment} will be dynamically bound to the existing parameter (Figure~\ref{fig:system-bind}B), so that when the $angle$ value changes, the angle of another line is also changed  (Figure~\ref{fig:system-bind}C).
This works as a dynamic constraint between multiple line segments.

\begin{demofigure}{h!}
\centering
\includegraphics[width=1\linewidth]{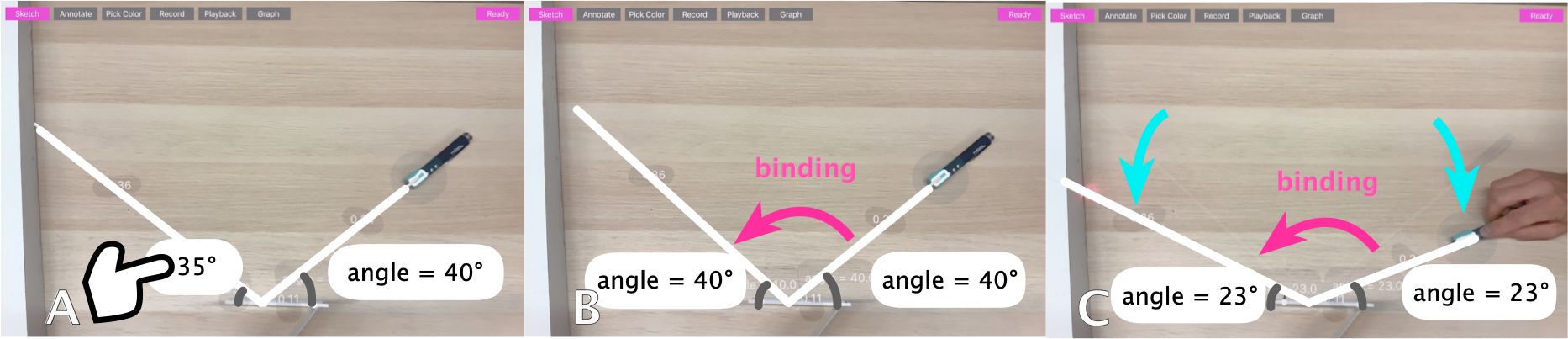}
\caption{Binding a variable: By using the same variable name for a static line segment (A-B), the variable can be bound to the existing dynamic line segment (C).}
~\label{fig:system-bind}
\end{demofigure}

Similarly, the user can define a constraint by typing a function of existing variable. 
For example, consider the user wants to draw the bisector of the angle formed by a {\it dynamic line segment}.
The user can first draw a line and an arc between the line and the base line (Figure~\ref{fig:system-function}A).
Then, the user taps the label and types $angle / 2$ (Figure~\ref{fig:system-function}B).
Then, the system can recognize the function to dynamically bind the parameter to be half of the existing parameter (Figure~\ref{fig:system-function}C).
In a similar manner, the user can also change the length or distance of the sketched elements with a function (e.g., the force of a spring in Figure~\ref{fig:system-overview}).
In our current implementation, the system can parse basic arithmetic operations, trigonometric functions, and polynomial expressions\nothing{, and quadratic functions\liyi{(June 2, 2020) I assume quadratic is a subset of polynomial unless it means something else.}}.

\begin{demofigure}{h!}
\centering
\includegraphics[width=1\linewidth]{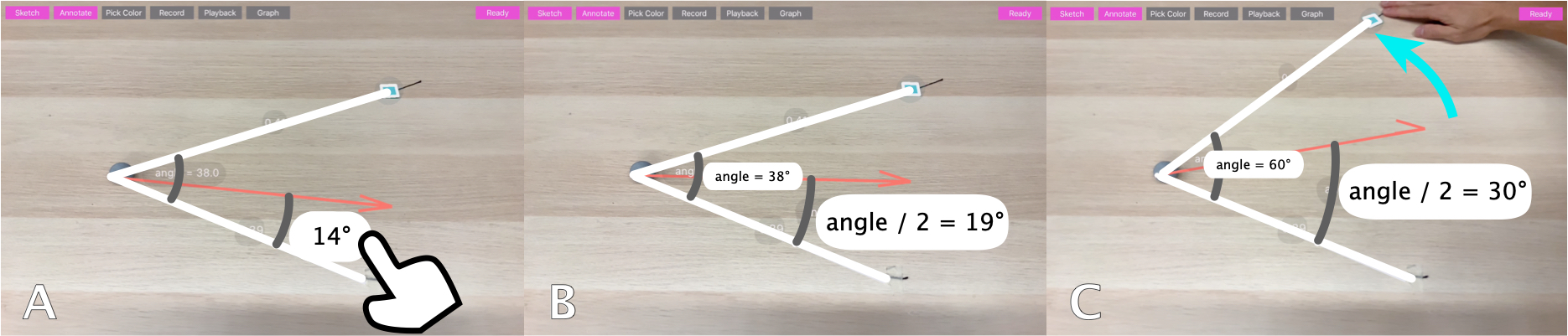}
\caption{Binding a function: The user can also bind a variable to a function e.g., angle / 2 (A-B), to make a dynamic constraint (C).}
~\label{fig:system-function}
\end{demofigure}

By default, if the user draws a {\it dynamic line segment} associated with a tracked object, the other end of the line is fixed to a certain point.
However, the user sometimes may want to draw a constant line segment that is attached to an object (e.g., an arrow to represent gravity or force).
In such cases, the user can tap annotation option for sketching, so that the distance and direction will be fixed when the tracked object moves (i.e., the endpoint is continuously calculated to maintain the initial relative position from the attached object).
This is useful to portray visual hints for hidden forces associated with the object\nothing{ (e.g., resultant force in Figure~\ref{fig:system-function})\liyi{(July 2, 2020) there is no resultant force there, or any figure I can find in the current paper draft.}}. 

\subsection{Visualizations}
\subsubsection{Visualize Motion with Responsive Graph Plots}
\tool{} can also make responsive visualization based on graph plotting of a parameter.
In the graph placing mode, the user can place a 2D graph and change its size by dragging and dropping onto the surface.
By default, the $x$-axis of the graph is time.
By binding an existing variable to the graph, it starts visualizing the time series data of the variable.

\begin{demofigure}{h!}
\centering
\includegraphics[width=1\linewidth]{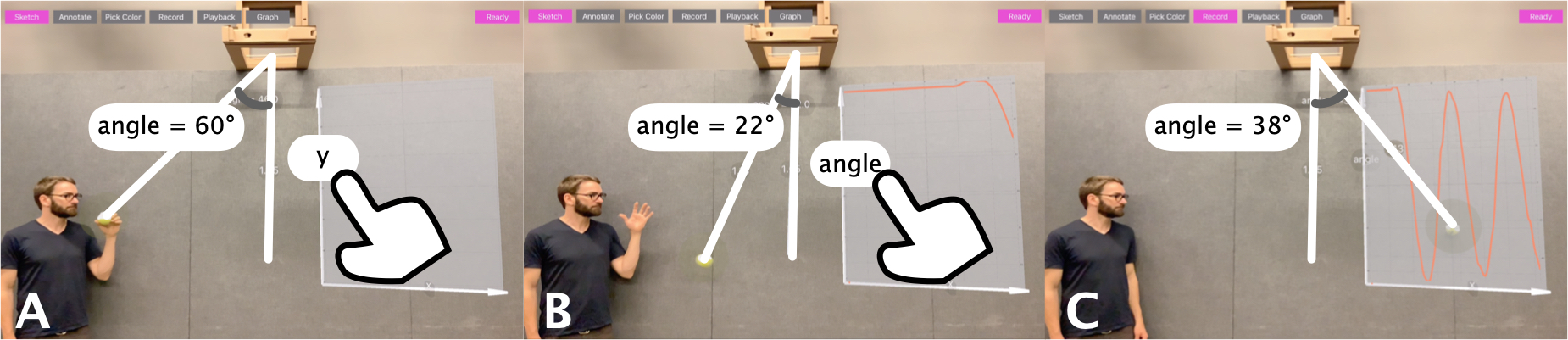}
\caption{Plotting a graph: By binding an axis label with an existing dynamic variable (A-B), the user can plot a real-time graph of the parameter (C).}
~\label{fig:system-plot}
\end{demofigure}

\begin{figure*}
\vspace{-0.2cm}
\centering
\includegraphics[width=\textwidth]{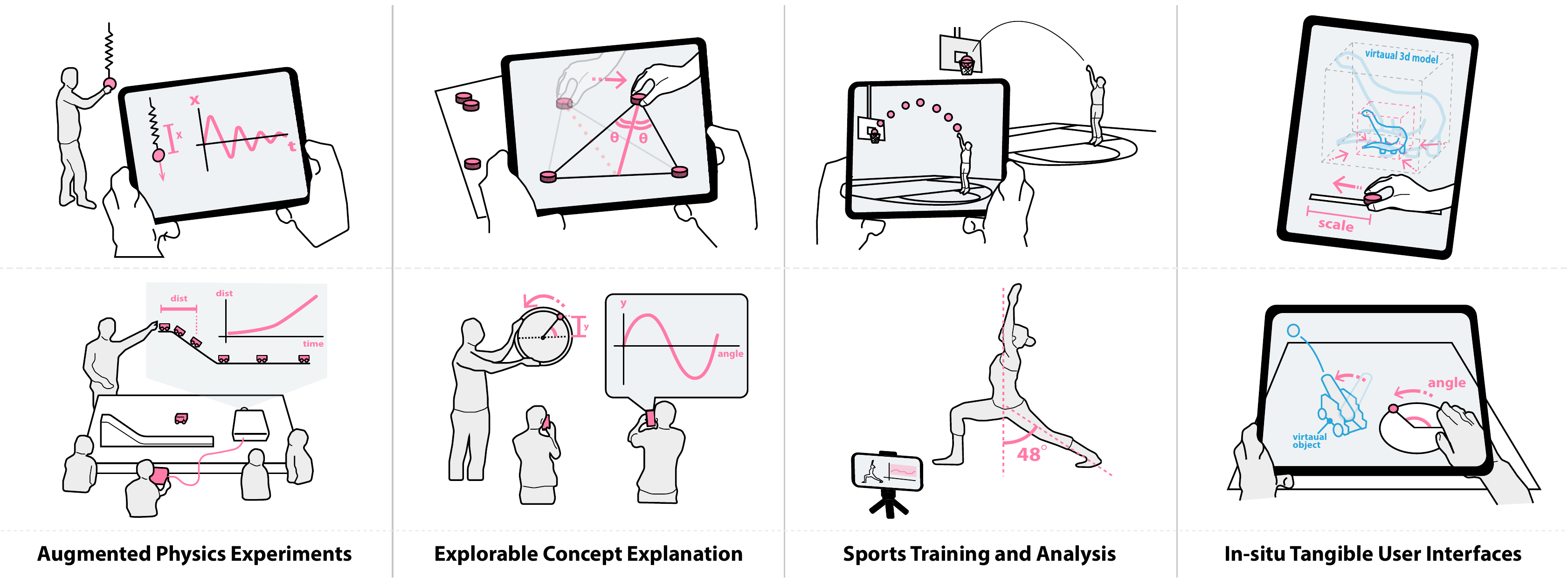}
\caption{Sketches of proposed application scenarios for \tool{}.}
~\label{fig:applications-scenarios}
\vspace{-0.6cm}
\end{figure*}

To bind the variable, the user can simply tap a label of the graph (Figure~\ref{fig:system-plot}A), and then, enter the variable the user defined.
For example, if the user binds the {\it angle} variable of the pendulum to the $y$-axis of the chart, the graph will dynamically plot the angle of the pendulum when it starts swinging (Figure~\ref{fig:system-plot}B-C).
By adding multiple variables separated with a comma (e.g., $a, b, c$), the user can also plot multiple parameters in the same graph.
The user can also change the $x$-axis from time to a variable by tapping the $x$-axis and entering a second variable.
For example, the user can define an angle and y distance of a point of a circle.
By binding x-axis as the angle and y-axis as the perpendicular length, the system can plot the relationship between the angle and corresponding sine value (e.g., Figure~\ref{fig:applications-scenarios}).

\begin{demofigure}{h!}
\centering
\includegraphics[width=1\linewidth]{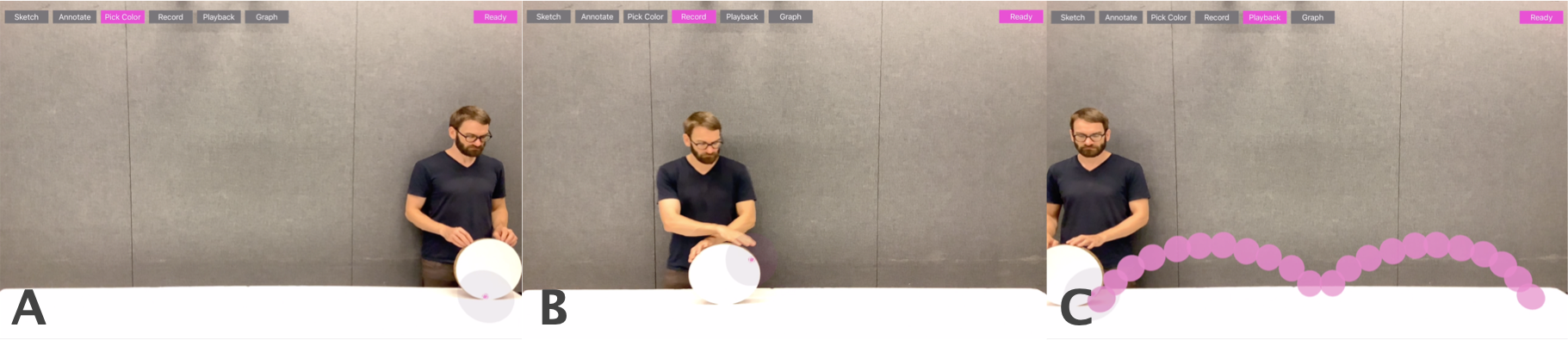}
\caption{Explanation of the cycloid curve: The user can visualize a point of the rolling circle to see its path is a cycloid curve.}
~\label{fig:system-record}
\end{demofigure}

\subsubsection{Visualize Motion by Recording and Replaying}
Finally, the system also supports recording a motion for analysis.
When the user taps the recording mode button, the system records the tracked objects' positions at each frame.
The recorded values are stored as an array of 3D positions.

When the user taps the playback button, the system visualizes the trajectory of the motion.
To render the time-lapse graphics, the system places 3D spheres of the same tracked color at each recorded 3D position in the world.
The user can also change the density of the visualization by changing the minimum distance between each object.
For example, if the minimum distance is zero, the system places spheres at all of the recorded positions in the scene (e.g., 200 objects for a 10 second recording at 20 FPS).
The default value of the minimum distance is set to 3 cm to better represent the motion and avoid slow performance due to rendering many objects.

\section{Application Scenarios}

\subsection{Augmented Physics Experiments}
In science education, a classroom experiment is an integral part of learning physics~\cite{wellington1998practical} because the real-world experiment provides students an opportunity to {\it connect} their knowledge with physical phenomena~\cite{hodson1996laboratory, mccomas1998role}.
\tool{} can become a powerful tool to support these experiments by leveraging real-time visualization capability.
Figure~\ref{fig:applications-pulley} illustrates how our tool can help students understand the pulley system.
In this scenario, by tracking the movement of two points (i.e., the point of the person grabs and the point of the load), the students can visualize the traveling distance of each point.
In this way, they can see the load distance movement is shorter than the distance the person pulls (Figure~\ref{fig:applications-pulley}B-C).

\begin{demofigure}{h!}
\centering
\includegraphics[width=1\linewidth]{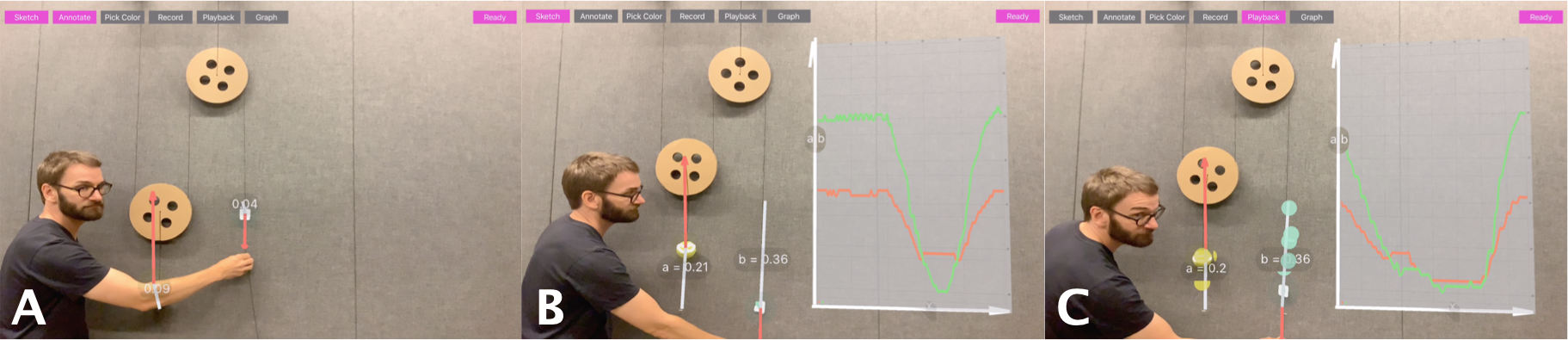}
\caption{Pulley system experiment: The user measures the distance of a hand movement and a load, and visualizes it with a graph (green: hand movement, orange: the \nothing{object}{load}).}
~\label{fig:applications-pulley}

\end{demofigure}

In the same way, our tool can help students visualize the motion of the pendulum experiment (Figure~\ref{fig:cover} and~\ref{fig:system-plot}) or behavior of the bouncing spring (Figure~\ref{fig:applications-scenarios}). 
In these examples, the user can not only visualize through a graph plot but also add responsive embedded annotations. 
For example, in Figure~\ref{fig:applications-pulley}, the user can add force vector represented with a red line that can dynamically move based on the position.
These annotated lines are helpful to let the user see the invisible phenomenon, such as a path of light reflection in Figure~\ref{fig:system-bind} or the resultant force of two pulled strings in Figure~\ref{\nothing{fig:system-function}fig:applications-pulley}.

\changes{Similarly, recording the trajectory of object motion can be helpful to understand physical phenomena.}
Figure~\ref{fig:applications-gravity} illustrates an experiment that demonstrates how the law of inertia applies to an object thrown by a moving person.
For this experiment, the user first tracks a ball and draws a ground line and vertical line perpendicular to the ground.
Then, the user starts tracking the distance of the ball from the ground.
By using the recording functionality, the trajectory of the ball is overlaid throughout time.
In this way, the student understands even if the person throws a ball in a vertical direction, the ball is continuously moving horizontally.
By capturing the height of the ball without lateral movement, the student also understands the vertical trajectory of the ball is the same, by overlaying the two graphs.

\begin{demofigure}{h!}
\centering
\includegraphics[width=1\linewidth]{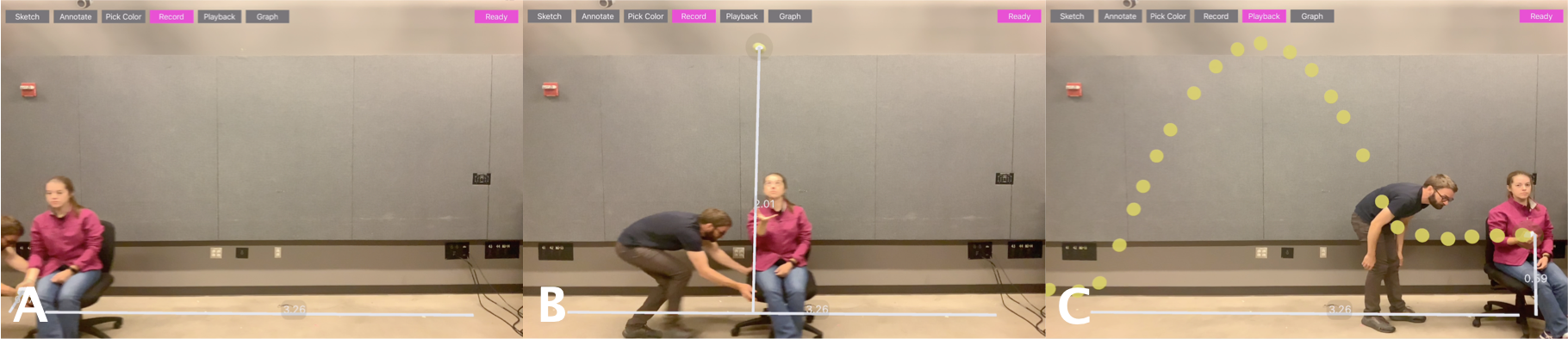}
\caption{Visualizing gravity and inertia force: The system tracks the height of the ball from the ground to visualize how the law of inertia applies to an object launched by a moving person.}
~\label{fig:applications-gravity}
\end{demofigure}

\subsection{Interactive and Explorable Concept Explanation}
\tool{} is also useful to help teachers explain concepts that may be difficult to understand with static graphs, and to let students explore them through tangible interactions.
Some examples are shown in Figure~\ref{fig:applications-scenarios} (Top: demonstrating an area of a triangle remains the same with horizontal movement; a bisector line of a triangle always intersect at the middle point. Bottom: showing how a sine curve is generated from plotting the angle and perpendicular distance of a rotating point.)

Figure~\ref{fig:applications-gear} shows an example that explains the relationship between gear ratio and angular motion.
By defining and tracking a rotating angle of each gear, the system continuously captures and visualizes the current angles of both gears, which explain the larger gear rotates slowly (e.g., half of the smaller gear).
The system is also useful to explain mathematical concepts.
For example, by tracking a point on a rolling circle along a surface, the system can show the trajectory is a cycloid curve (Figure~\ref{fig:system-record}).
Traditionally, these interactive concept explanations are done through video production~\cite{hidden, byjus} or programming~\cite{explorables, concept-visualization}, but with \tool{}, it is easier to explain and explore through real-world objects.

\begin{demofigure}{h!}
\centering
\includegraphics[width=1\linewidth]{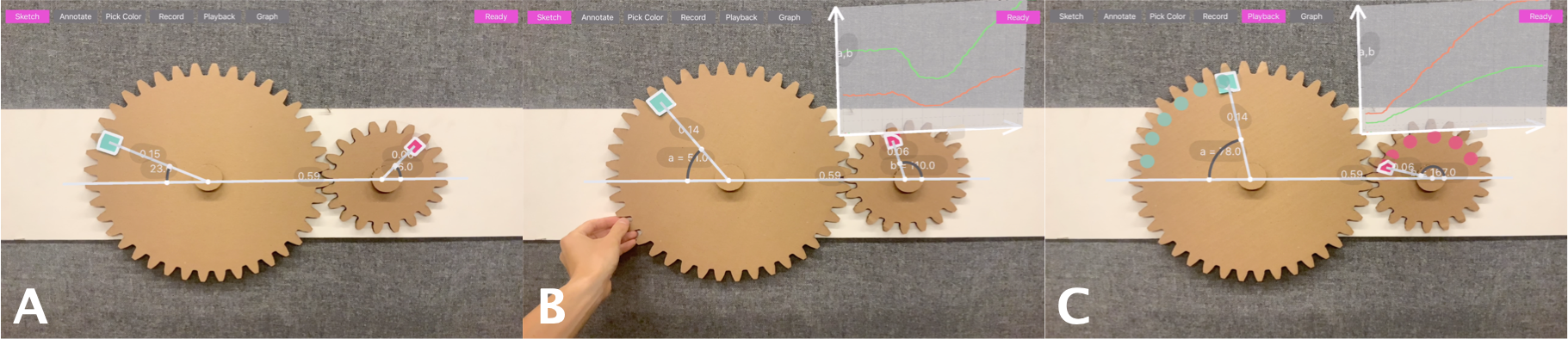}
\caption{Explanation of gear ratios: The teacher explains the angular motion of different sized gears.}
~\label{fig:applications-gear}
\end{demofigure}

For educational applications, we can think of the following three setups of how students and teachers can use our tool:
\begin{description}[itemsep=-1mm]
\item [1. Classroom presentation:]
In this case, a teacher or an assistant sketches and visualizes the motion, which can be shared through a connected large display or projector, so that the students can see and understand.
\item [2. Collaborative learning:]
In this case, students can form a group and interact with their own devices.
Since mobile AR is accessible for almost all smartphones, every student should be able to play by themselves, which can lead to an interesting exploration and discoveries. 
\item [3. Remote learning:]
In this case, a teacher only records the video of the experiment, then share the recorded video with the students.
The student can download and visualize with their own app, so that it provides a fun and interactive experiment to support online and remote learning.
\end{description}

\subsection{Improvised Visualization for Sports and Exercises}

\tool{} could be also useful to analyze and visualize the motion for sports training and exercises because they often employ the physical movements. 
For example, sports practices, like a golf shot, baseball pitching, and basketball shooting, would be an interesting example to visualize the trajectory of the ball (Figure~\ref{fig:applications-scenarios}).
Similar to the previous example of showing the trajectory of a ball (Figure~\ref{fig:applications-gravity}), it is useful to quickly see the path through stroboscopic effects. 
In addition to showing the trajectory, the system can also capture and compare multiple attempts to let the user analyze what works and what does not.

\begin{demofigure}{h!}
\centering
\includegraphics[width=1\linewidth]{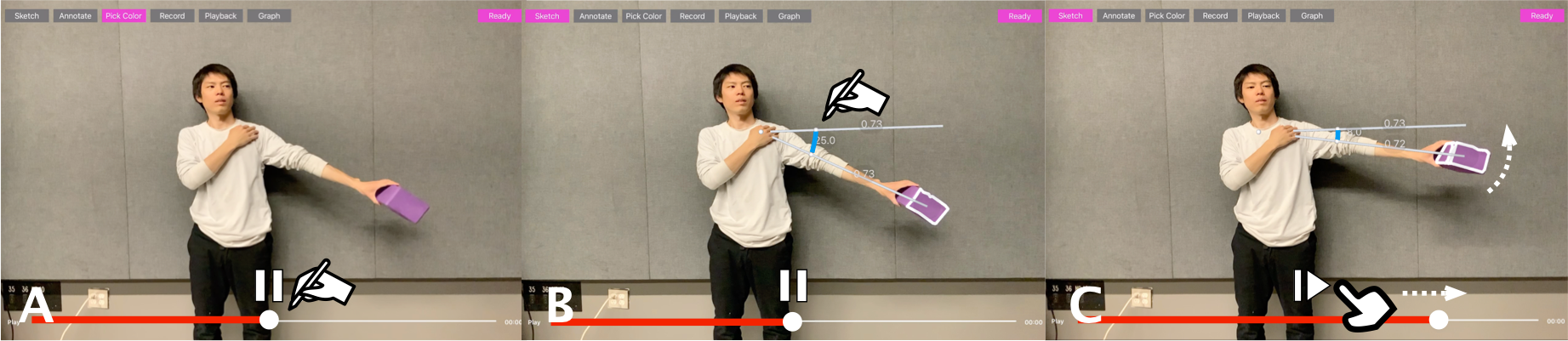}
\caption{Rehabilitation support: The recording feature aids the user's self-training. The user first records the motion, then interactively visualizes the posture by controlling the recorded video.}
~\label{fig:applications-rehab}

\end{demofigure}

Also, many sports and exercises may require a proper form and posture. 
For example, in baseball batting, golf shot, and a tennis swing, a player's form, such as a body angle, can be important, which the tool can help visualize through sketched guided lines.
These features could be particularly useful for exercise instructions. 
For example, in yoga practice, bike riding, or weight lifting training, there are recommended angles to be followed to maximize the performance benefits.
In these cases, the improvisational measurement of the posture can help the user to see and check if correctly done (\Cref{fig:applications-rehab,fig:applications-stretch}). 

\begin{demofigure}{h!}
\centering
\includegraphics[width=1\linewidth]{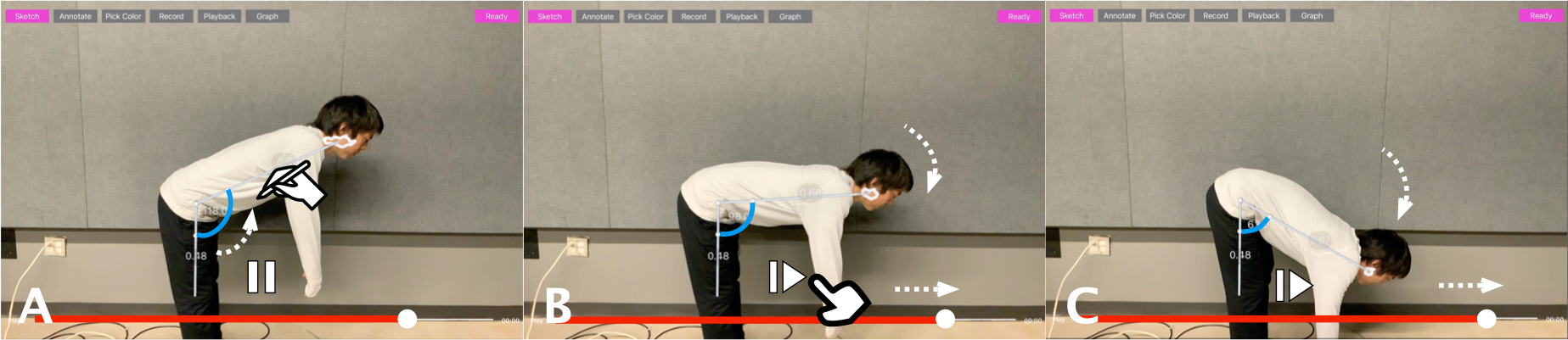}
\caption{Exercise support: The user measures an angle of body joints for yoga, stretching exercises, and weight lift training, which helps the user check and see if they follow the correct form.}
~\label{fig:applications-stretch}

\end{demofigure}

These scenarios can work with co-located or remote instruction where an instructor measures and gives feedback in real-time.
It also works as a self-supporting tool in which the user can capture and later analyze it by leveraging the video recording feature.
This feature is also useful for rehabilitation assistant. 
By measuring and visualizing it, the patients can better understand how they should fix it. 
The body joint tracking supported by ARKit is particularly useful for these scenarios as it provides a more handy way, compared to attaching distinct colored markers on the body.

\subsection{In-situ Tangible User Interfaces}

The parameterized value can be used for many different purposes to enable responsive visual outputs. 
So far, we have mostly focused on animation of the simple basic geometry (e.g., line segments, arc) or build-in visualization outputs (e.g., graph plot). 
However, the dynamic parameter value can be also used for other outputs via binding, as we discussed in the previous sections (e.g., pre-defined vs user-defined section).

\begin{demofigure}{h!}
\centering
\includegraphics[width=1\linewidth]{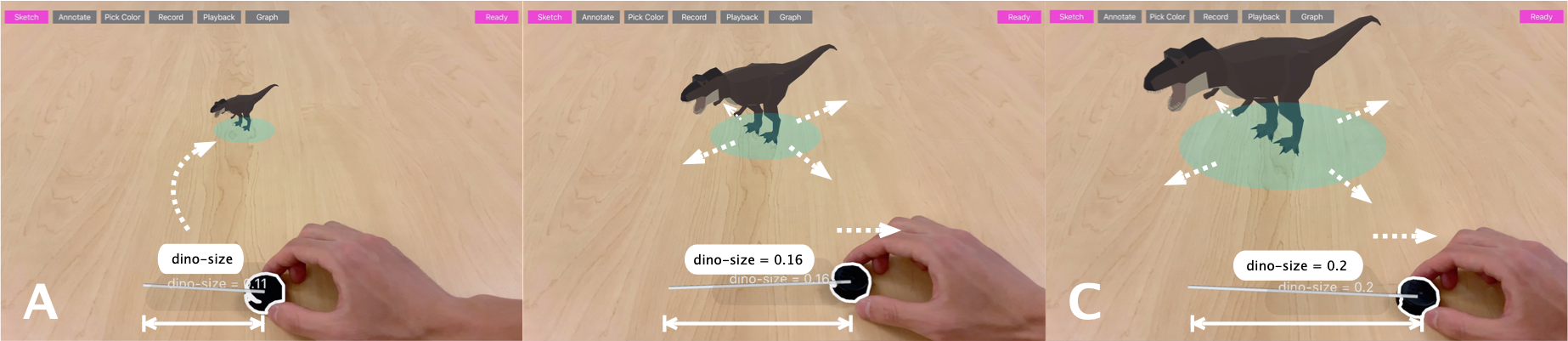}
\caption{In-situ creation of a tangible UI slider: The user binds the object distance to the scale of the virtual 3D model, so that the model size changes as the object moves.}
~\label{fig:applications-dino}
\end{demofigure}

\begin{demofigure}{b!}
\centering
\includegraphics[width=1\linewidth]{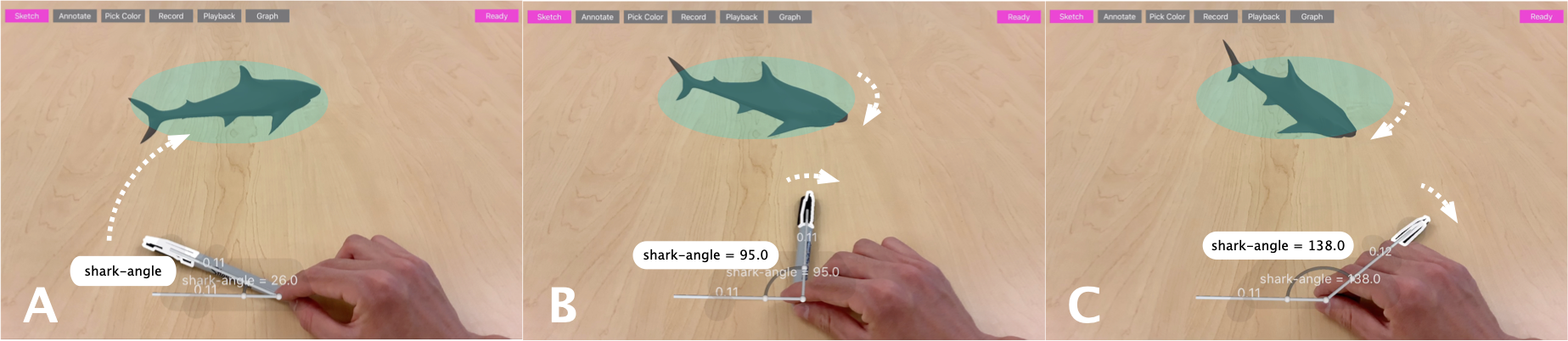}
\caption{By mapping the defined variable to an angle property of the virtual object, the user can quickly craft an improvised tangible dial to rotate the virtual object.}
~\label{fig:applications-shark}

\end{demofigure}

One promising application domain of this is to use these dynamic parameter values as an input of changing a property of existing virtual objects.
For example, if one can connect a parameter value to a size property of a virtual 3D model, then the size of the model can dynamically change when the value changes. 
This use case is particularly useful for in-situ creation of tangible controllers (Figure~\ref{fig:applications-scenarios}). 
For example, a colored token can become a tangible slider to change the size of the object (Figure~\ref{fig:applications-dino}).
The system could bind these values based on a proper naming rule (e.g., {\it ``dino-size''} in Figure~\ref{fig:applications-dino}, {\it ``shark-angle''} in Figure~\ref{fig:applications-shark} , and {\it ``tree-num''} in Figure~\ref{fig:applications-tree}).

In addition to sliders, the user can also quickly craft various, improvised controller such as toggle buttons, joysticks, 2D sliders, and dial sliders, by appropriately defining the parameter (e.g., length, position, and angle), and constraints based on expression (e.g., output = (length > 10) to make it binary for the toggle button).
Traditionally, making such interactive physical controllers require a dedicated electro-mechanical system (e.g., Arduino) or a well-prepared kit (e.g., Nintendo Labo), but our approach could provide a more handy way to create an in-situ controller for quick use or prototyping, similar to CV-based controllers like ARcadia~\cite{kelly2018arcadia}.

\begin{demofigure}{h!}
\centering
\includegraphics[width=1\linewidth]{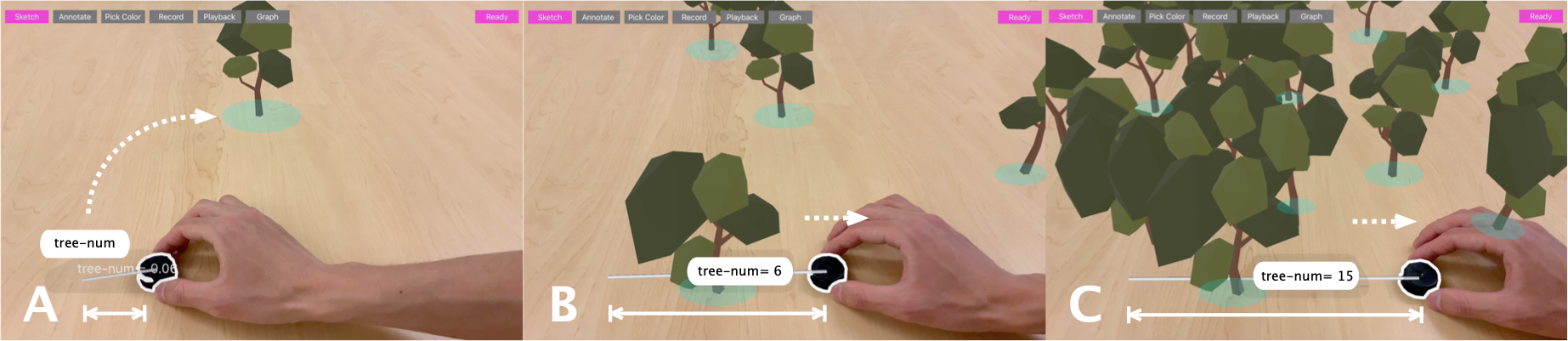}
\caption{Tracked object parameters can be mapped to a program property through a pre-defined variable name. In this case, the number of virtual trees is controlled by a tangible slider.}
~\label{fig:applications-tree}
\end{demofigure}




\section{Preliminary Evaluation}

\subsection{Usability Study}
\subsubsection{Method}
\changes{We validated our system with two preliminary studies: usability study and expert review. The goal of the first study is to evaluate the usability of our prototype and to identify limitations or opportunities for future improvements.}
To do so, we recruited six participants (3 male, 3 female, ages 23-33) from our university (3 computer science, 3 engineering).
All sessions took place in a research lab, and consisted of two steps. In the first step, the interviewer demonstrates all of the basic functionalities of our prototype system by going through the light reflection example (Figure~\ref{fig:system-tracking}-\ref{fig:system-bind}).
Then, the participants were asked to perform example tasks without the author's help.
\changes{Due to time constraints, we chose two examples (visualizing the motion of a pendulum and the trajectory of a thrown ball) as evaluation tasks.}
After the session, we asked the participants to give feedback about the interface and interactions with an online questionnaire form.
Sessions lasted approximately 30--45 minutes and participants were compensated USD 10.

\subsubsection{Results}
All our participants were able to complete the given task without any assistance.
In general, participants responded positively to the usability, applications, and unique affordances of \tool{}. 
The participants responded that the interactions were intuitive (P1, P2), easy to learn (P5), and enjoyable (P3).
``\textit{The interface is simple and highly responsive. Sketching lines and angles, the color-detection, graphing, and playback all felt natural and useful}'' (P4). 
Participants' average rating of the workflow is 5.83 out of 7 (min 5, $\sigma = 0.68$), and the average rating for engagement is 6.83 out of 7 (min 5, $\sigma = 0.37$).

All the participants saw the potential benefits of \tool{} for classroom teachings, student engagements, enhanced understanding, and in-class work exercises.
In particular, participants felt that \tool{} provides an ``\textit{interactive and engaging experience to the students to visually see and involve in creating the abstractions of concepts }'' (P3).
Participants also stated their desire to use such AR interactions in a variety of applications beyond teaching, including engineering visualization (P1), annotating of worksites (P3), sports analysis (P5), interactive design discussions (P2), and remote consultation for medical applications (P2). 
``\textit{This system can be used for engineering prototyping, where low-fidelity physical models are annotated by diagrammatic elements}'' (P2).

\subsubsection{Suggestions for Future Improvements}
While the study participants were generally satisfied with \tool{}, they also gave suggestions for future improvements. 
For example, the lack of undo and delete operations often hindered their experience (P1, P2, P3, P6).
Also, snapping features like~\cite{nuernberger2016snaptoreality} \changes{were} also suggested by the participants to make the drawing easier (P2).

The video recording feature was highly appreciated, but the participants also wanted to analyze pre-existing videos available on YouTube, as there are a lot of interesting resources online (P3, P4).
Although our current prototype does not support this feature, machine learning based depth reconstruction of 2D videos~\cite{li2019learning} could make this possible.
The participants also suggested it would be easier to select variables without actually typing them (P3). For example, when tapping the label, it would be better to show the list of pre-defined variables for easier selection.
The participants also suggested us to enable the same interactions for spatial augmented reality (e.g., projection mapping), so that multiple users can manipulate the object and share the same experience, which would be more suitable for classroom examples (P3).
Finally, they wanted to experience this interface in HMD-based augmented or mixed reality (e.g., Microsoft Hololens), so that the experience can become more immerse. 

\subsection{Expert Review}
\subsubsection{Method}
\changes{We also conducted an expert review to gain in-depth feedback from the application scenarios.}
We recruited six experienced classroom instructors (E1-E6) and sports and yoga instructors (E7-E8) to get expert feedback about how \tool{} can be integrated into their future practice.
E1-E6 have at least 8 years of experience in teaching math- and physics-related classes at the college and high school level, E7 has 2 years of experience in teaching yoga, and E8 has 4 years of experience as a baseball coach.
The interviews took place in face-to-face (E1-E4) or remote discussion settings (E5-E8).
We first demonstrated the system with a simple example and showed some of the videos to help participants understand the feature. 
Then, we conducted an in-depth open-ended discussion about our approach and use scenarios.
The interview lasted approximately one hour for each expert.

\subsubsection{Insights and Feedback}
All the participants were very excited about the possibility of this tool.
They perceived using a commodity device as a powerful way to help students learn physics (E2, E4, E5).
``\textit{I absolutely like the idea of giving them an experiment as homework when you don't need a lab setting, as students don't have that kind of stuff [devices for measuring experiments] at home}'' (E2).

Teachers (E3, E5) also mentioned the benefits of using real-world experiments, as compared to computer simulations.
E3 saw the possibility of teaching non-intuitive mathematical concepts by enabling students to try them out themselves, referring to an example that connecting the midpoints of an arbitrary quadrilateral forms a parallelogram~\cite{borning1981programming}.
``\textit{Say, if you have four colored objects and connect the midpoints of these four lines, then you can interactively see this theorem works by manipulating the objects}'' (E3).
Such an interactive experience can provide more compelling impressions for students (E1, E3) and allow students to gain intuition for math (E2).

One instructor (E1) shared a suggestion for improvement:
``\textit{Currently, the tool can only visualize the experimental data, but it would become more interesting if we can also show and compare the difference between the experimental and theoretical (simulated) data}'' (E1).
They mentioned that in this way, for example, the tool could show how the ball falls in a different condition, such as on the Earth, Moon, and Mars.

The participants were also excited about the applications for sports and exercise analysis (E7-E8).
``\textit{Novice learners often struggle to keep an appropriate posture, so this would be useful for them to check}'' (E7).
Currently, they often use a mirror to check the posture, but visualizing the body's skeleton through sketching would be useful to analyze the posture (E7).
While some felt the sketching interactions might be tedious for everyday use, particularly for expert users like themselves (E7-E8), they also found the sketching interactions interesting as a feature to keep end users engaged (E7-E8). 

\section{Limitations and Future Work}
\subsubsection{More General and Robust Object Tracking}
In our current implementation, we use simple color matching for tracking. However, this tracking method becomes unstable if the object does not have a solid color or if objects of similar color exist in the scene. We have also tested with other tracking methods leveraging recent algorithms (e.g., YOLO~\cite{redmon2016you}, Faster R-CNN~\cite{ren2015faster}, Mask R-CNN~\cite{he2017mask}). Although these algorithms are more robust under certain circumstances, some of them are slow on a mobile phone when tested (0.1-5 FPS) and require a training period.
However, we believe the continuous improvement in real-time computer vision algorithms along with advances in camera hardware (e.g., LiDAR of the iPad Pro 2019) promises to provide more robust, general, and fast tracking methods in the future.

\subsubsection{Spatial Interactions for Augmented Storytelling}
In mobile AR settings, all interactions are limited to the 2D tablet screen.
In the future, we hope to explore how the proposed method can be applied to other configurations (e.g., HMD-based mixed reality or projection mapping) by leveraging spatial and whole-body interactions. Rather than sketching on the mobile device screen, users could for instance sketch directly onto surfaces in the world.
Alternative configurations could allow the instructor to directly manipulate and visualize without requiring an assistant.
This promises other exciting applications such as interactive storytelling, interaction prototyping, and augmented presentations~\cite{saquib2019interactive}.

\subsubsection{Parameterization and Visualization beyond Physical Motion}
\changes{Currently, our prototype only supports simple constraints, which limits the possible visualization.
By leveraging the expressive constraints  demonstrated in existing tools (e.g., Sketch it, Make it~\cite{johnson2012sketch}, Shapr3D~\cite{shapr3d}), we can further expand our idea to more expressive animation and visualizations.}
Also, our current parameterizations are limited to spatial parameters (e.g., length, angle, area), but there is a broader space of potential real-world parameterizations.
Other possibilities include other visual variables (e.g., size, shape, orientation, color, texture, object recognition)~\cite{Bertin:1983:SG}.
Also, the real-time visualizations can be driven by existing sensors or other commodity devices~\cite{google-journal}, e.g., plotting voltage by reading a potentiometer value, visualizing a real-world magnetic field by virtually duplicating a magnetic compass, or visualizing acoustic resonance by spatially recording from a mobile phone's microphone.
Expanding the scope of input parameters, output graphics (e.g., 3D models), and mapping mechanisms would enable diverse applications to leverage the benefits of embedded and responsive graphics through \changes{dynamic sketching}.
From a design perspective, we would like to further explore the combination of pre-defined procedures to add more complex (procedural) behaviors to user drawn sketches, parameterizations, and physical objects.

\changes{
\subsubsection{Sketching with HMD-based Augmented and Mixed Reality}
While we chose mobile AR setup for our prototype primarily because of its wide availability, we also look forward to extending our interactions to head-mounted display (HMD) based setups like Microsoft Hololens to improve immersion. Particularly, in-situ tangible interaction could become a good alternative or complement to current mid-air gestural interactions with HMDs.
\tool{} relies on the mobile device screen for sketching. We expect that this interaction could be adopted to HMD, where the user instead sketches on an existing physical surface (e.g., tabletop) or use the tablet to sketch, then see the sketched elements through the HMD, similar to SymbiosisSketch~\cite{arora:2018:symbiosissketch} and PintAR~\cite{gasques2019you}.
}

\subsubsection{Deployment to Actual Environments}
\changes{Our preliminary evaluation is limited in scale and further in-depth evaluation would be required.}
For future work, we intend to deploy our tool into classroom settings (introductory physics course) for an in-depth and long-term study. This would help us to gain insights about its impact in students' engagement and learning activities. 





\section{Conclusion}
We presented \tool{}, a tool that allows users to sketch interactive graphics in AR. 
By interactively binding sketched elements with physical objects, these embedded graphics can dynamically \nothing{change}\changes{respond} when the physical objects move.
This paper described the system implementation of \tool{} and demonstrated the flexibility of the tool through a range of possible applications.
We hope this paper opens up new opportunities for embedded and responsive sketching and inspires the HCI community to further explore such interactions to realize the full potential of AR and MR as a dynamic, computational medium.

\section{Acknowledgements}
We thank our anonymous reviewers for their suggestions on improving this paper. 
We also thank the participants of our usability study and expert review.
This work is partly supported by an Adobe Gift Funding and the Nakajima Foundation scholarship.

\balance
\bibliographystyle{SIGCHI-Reference-Format}
\bibliography{references,misc}

\end{document}